\documentclass[aps,pra,twocolumn,amsmath,amssymb,showpacs,nofootinbib,longbibliography]{revtex4-1}

\newcommand{\Jnature}{Nature (London)}
\newcommand{\Jnatphys}{Nat. Phys.}

\newcommand{\Jscience}{Science}

\newcommand{\Jprx}{Phys. Rev. X}
\newcommand{\Jprl}{Phys. Rev. Lett.}
\newcommand{\Jpr}{Phys. Rev.}
\newcommand{\Jpra}{Phys. Rev. A}
\newcommand{\Jprb}{Phys. Rev. B}

\newcommand{\Jrmp}{Rev. Mod. Phys.}

\newcommand{\crasphy}{C. R. Phys.}

\newcommand{\Jadvphys}{Adv. Phys.}

\newcommand{\JAnnualRevCondMat}{Annual Rev. Cond. Mat. Phys.}

\newcommand{\Jprogthphyssup}{Prog. Theor. Phys., Suppl.}

\newcommand{\Jphysicascripta}{Phys. Scr.}

\usepackage[english]{babel}
\usepackage{latexsym}
\usepackage{graphics}
\usepackage{subfigure}
\usepackage{epsfig}
\usepackage{color}
\usepackage{hyperref}
\usepackage{braket} 
\usepackage[T1]{fontenc}
\usepackage[latin9]{inputenc}
\usepackage{amstext}
\usepackage{amssymb}
\usepackage{graphicx}
\usepackage{esint}
\usepackage{braket}
\usepackage{babel}
\usepackage{amsmath}
\hypersetup{
colorlinks=true,
citecolor=blue,
linkcolor=red,
urlcolor=black
}

\usepackage{times}

\newcommand\avg[1]{\langle #1 \rangle}
\renewcommand{\t}[1]{\text{#1}}
\renewcommand{\d}{\mathrm{d}}
\renewcommand{\vec}[1]{\mathbf{#1}}
\newcommand{\mc}[1]{\mathcal{#1}}
\newcommand{\e}{\textrm{e}}
\newcommand{\ie}{i.e.}
\newcommand{\eg}{{e.g.}}

\definecolor{orange}{rgb}{0.85,0.5,0.2}


\begin{document}

\newcommand{\ttitle}{Unraveling the excitation spectrum of many-body systems from quantum quenches}
\title{\ttitle}

\author{Louis Villa}
\affiliation{CPHT, CNRS, Institut Polytechnique de Paris, Route de Saclay 91128 Palaiseau, France}

\author{Julien Despres}
\affiliation{CPHT, CNRS, Institut Polytechnique de Paris, Route de Saclay 91128 Palaiseau, France}

\author{Laurent Sanchez-Palencia}
\affiliation{CPHT, CNRS, Institut Polytechnique de Paris, Route de Saclay 91128 Palaiseau, France}
\date{\today}

\begin{abstract}
Quenches are now routinely used in synthetic quantum systems to study a variety of fundamental effects, including ergodicity breaking, light-cone-like spreading of information, and dynamical phase transitions.
It was shown recently that the dynamics of equal-time correlators may be related to ground-state phase transitions and some properties of the system excitations.
Here, we show that the full low-lying excitation spectrum of a generic many-body quantum system can be extracted from
the after-quench dynamics of equal-time correlators.
We demonstrate it for a variety of one-dimensional lattice models amenable to exact numerical calculations,
including Bose and spin models, with short- or long-range interactions.
The approach also applies to higher dimensions, correlated fermions, and continuous models.
We argue that it provides an alternative approach to standard pump-probe spectroscopic methods and discuss its advantages.
\end{abstract}

\maketitle
\section{Introduction}
The properties of the low-lying excitations on top of the ground state are an essential feature of a quantum many-body system.
They govern a variety of fundamental phenomena, from
electronic conductivity and superfluidity to
quasi-long-range order in low dimensions~\cite{negele1988,mahan2000,giamarchi2004}.
For a wide range of correlated systems, they are efficiently described by the notion of quasiparticles (including phonons, plasmons, spinons, magnons, Bogoliubov particle-hole pairs, and doublon-holon pairs).
In practice, the elementary excitations of a system at equilibrium are commonly probed through the spectral representation of an unequal-time correlator (UTC), for instance, the spectral function or the dynamical structure factor~\cite{bruus2004,rickayzen2013}.
Yet, the analytical or numerical derivation of the latter remains a formidable task in strongly correlated systems,
even for integrable ones~\cite{caux2006dynamical,pippan2009excitation,ejima2012dynamic}.
In experiments, they arise from tedious pump-probe spectroscopic techniques,
such as angle-resolved photoemission spectroscopy (ARPES),
inelastic neutron or x-ray Raman scattering,
and two-photon Bragg spectroscopy~\cite{furrer2009,damascelli2004,stewart2008using,stenger1999bragg,ozeri2005colloquium,clement2009exploring,meinert2015probing}.

The dramatic progress made in recent years on the time-resolved control and the out-of-equilibrium dynamics of isolated quantum systems~\cite{polkovnikov2011,eisert2015,langen2015,lewenstein2007,bloch2008, NaturePhysicsInsight2012cirac, *NaturePhysicsInsight2012bloch,*NaturePhysicsInsight2012blatt,*NaturePhysicsInsight2012aspuru-guzik,*NaturePhysicsInsight2012houck,gross2017,lsp2018,*tarruell2018,*aidelsburger2018,*lebreuilly2018,*LeHur2018,*bell2018,*alet2018}
allows us to reconsider these issues from the perspective of quench dynamics.
A large body of work is devoted to understanding fundamental effects, including the onset of thermalization and its breaking, dynamical phase transitions, and the emergence of causality in information spreading.
Out-of-equilibrium dynamics may also be considered in connection to equilibrium properties~\cite{lang2018,halimeh2018quasiparticle,hashizume2018dynamical},
and it was recently proposed to probe ground-state phase transitions using quenches~\cite{prosen2000,roy2017,heyl2018,titum2019,daug2019detection}.
It is then natural to ask whether information about the system excitations can be extracted from quenches.
For instance, it has long been recognized that the Lieb-Robinson bound for information spreading in short-range lattice models may be related to the maximum group velocity~\cite{lieb1972finite,calabrese2005evolution,calabrese2006time}.
More recently, it has been shown that the structure of correlations in the vicinity of the causal edge can be related to
basic properties of the elementary excitations, including characteristic velocities, dynamical exponents, and gaps~\cite{cevolani2018universal,despres2019twofold}.

In this paper, we show that the full low-lying excitation spectrum of a correlated quantum system can be extracted from equal-time correlators (ETC) following a global quench.
We develop a general framework for unravelling excitation spectra and measure them experimentally.
It generalizes previous results using power spectrum analysis of density ripples in one-dimensional quasicondensates~\cite{schemmer2018} and spin correlations in two-dimensional models with flat bands~\cite{menu2018}.
We introduce the \textit{quench spectral function} (QSF) and show that it yields the quasiparticle dispersion relation, irrespective of the system dimension, particle statistics, range of interactions, and the discrete or continuous nature of the model.
We illustrate this on one-dimensional models by computing the exact QSF using time-dependent matrix product state calculations.
We first use the Bose-Hubbard model as a benchmark in both the Mott insulator and mean-field superfluid phases, and recover known analytical dispersion relations.
In the strongly interacting superfluid regime, where no exact result is known, we show that the QSF exhibits a continuum of excitations, which we interpret by devising an approximate Bethe ansatz method.
Further, we extend our results to other quantum models,
using the long-range transverse Ising model as a paradigmatic example.
We argue that the QSF approach provides an accurate method to probe the excitation spectrum of correlated quantum models and discuss its advantages compared to standard pump-probe spectroscopy.

\section{Quench spectral function}
We start with the system in some initial state, described by the density matrix $\hat{\rho}_\textrm{i}$, and induce out-of-equilibrium dynamics by performing a quench at time $t=0$. The dynamics is then governed by the Hamiltonian $\hat{H}$, such that  $\hat{\rho}_\textrm{i}$ is non-stationary ($[\hat{\rho}_\textrm{i}, \hat{H}] \neq 0$). We consider the ETC
\begin{equation}
\label{eq:observables}
G(\vec{R},t)=\avg{\hat{O}_{1}^{\dagger}(\vec{R},t)\hat{O}_{2}(\vec{0},t)},
\end{equation} 
where $\hat{O}_{j}(\vec{R},t)$ is a local operator at position $\vec{R}$ and time $t$,
and $\avg{\hat{X}}=\text{Tr}(\hat{\rho}_\textrm{i}\hat{X})$ is the average over the initial state.
For a translation invariant system, its spectral representation (aka quench spectral function), reads as (see Appendix~\ref{sec:SM_derivation_QSF})
\begin{equation}
\label{eq:QSF_spectral_form}
\begin{split}
G&(\vec{k},\omega)\propto\sum\limits_{n,n',m}\rho_\textrm{i}^{n' n}\bra{n}\hat{O}_{1}^{\dagger}\ket{m}\bra{m}\hat{O}_{2}\ket{n'}\times\\
&\delta(E_{n}-E_{n'}-\omega)\delta(\vec{P}_{m}-\vec{P}_{n'}-\vec{k})\delta(\vec{P}_{n}-\vec{P}_{n'}).
\end{split}
\end{equation}
The kets $\ket{n}$ have a well defined momentum $\vec{P}_{n}$ and span an eigenbasis of $\hat{H}$,
$\hat{O}_{j}=\hat{O}_{j}(\vec{0},0)$ is the operator at the origin of space and time,
and we set $\hbar=1$.
The most important feature of Eq.~(\ref{eq:QSF_spectral_form}) is the emergence of the dynamical selection rule $E_{n}=E_{n'}+\omega$. This applies regardless of the nature of the eigenstates, provided that the operators $\hat{O}_1$ and $\hat{O}_2$ couple the states $\vert n \rangle$ and $\vert n' \rangle$.
It permits us to identify the transition energies $E_{n}-E_{n'}$ to the resonance frequencies $\omega$, as in standard spectroscopy.

It is worth noting, however, that the QSF differs from the dynamical structure factor associated to the operators $\hat{O}_{1}$ and $\hat{O}_{2}$, which is measured by pump-probe spectroscopic methods.
The fundamental difference is that, here, $\hat{\rho}_\textrm{i}$ and $\hat{H}$ cannot be diagonalized simultaneously.
The density matrix therefore contains nonvanishing coherence (off-diagonal) terms,
$\rho_\textrm{i}^{n'n} \neq 0$ with $n' \neq n$.
The latter create the dynamical selection rule in Eq.~(\ref{eq:QSF_spectral_form}).
This is an essential consequence of the fact that the state being probed is out of equilibrium.
In contrast, the dynamical structure factor probes an equilibrium state and the dynamical selection rule appears only if one considers an UTC, \ie\ $G(\vec{R},t)=\avg{\hat{O}_{1}^{\dagger}(\vec{R},t)\hat{O}_{2}(\vec{0},t')}$ with $t \neq t'$ (see Appendix~\ref{sec:SM_derivation_DSF}).
Another important difference is that, in contrast to dynamical structure factors, the QSF can be measured using global, homogeneous, quench experiments. The latter are now routinely performed in atomic, molecular, and optical (AMO) physics and may considerably simplify the spectroscopy of many-body systems (see below).

Let us now assume that the initial state is close to the ground state $\ket{0}$, so that $\rho_\textrm{i}^{n'n}$ is non-negligible only when either $\ket{n}$ or $\ket{n'}$ is $\ket{0}$.
This condition is fulfilled for weak enough quenches. Focusing on the positive frequency sector, it sets $\ket{n'}=\ket{0}$. 
Assuming that $\hat{O}_{j}$
is a weakly coupling operator,
the intermediate states $\ket{m}$ in Eq.~(\ref{eq:QSF_spectral_form}) can be restricted to single quasiparticle excitations (see Appendix~\ref{sec:SM_derivation_QSF}).
The second selection rule in Eq.~(\ref{eq:QSF_spectral_form}) imposes $\ket{m}=\ket{\vec{k}}$,
\ie\ a quasiparticle of momentum $\vec{k}$.
Finally, the third selection rule imposes $\vec{P}_n=\vec{0}$.
The lowest-excited states that meet this criterion are composed of pairs of quasiparticles with opposite momenta,
$\ket{\vec{k},-\vec{k}}$.
For each momentum $\vec{k}$, the QSF thus produces a resonance at the frequency $\omega=2E_{\vec{k}}$,
hence providing the excitation dispersion relation.

\section{Benchmarking}
We now benchmark our approach against exact results, using first the one-dimensional Bose-Hubbard model (BHm),
\begin{equation}\label{BHm}
\hat{H}=-J \sum_{R} \left(\hat{a}^{\dagger}_{R}\hat{a}_{R+1}+\t{H.c.}\right)+\frac{U}{2}\sum_{R} \hat{n}_{R}(\hat{n}_{R}-1),
\end{equation}
whose in and out-of-equilibrium properties have been extensively studied~\cite{fisher1989boson,cazalilla2011one,schutzhold2006sweeping,fischer2008bogoliubov,trotzky2012probing,cheneau2012light,barmettler2012propagation,carleo2014light,villa2018cavity}.
In brief the BHm describes interacting bosons on a lattice,
characterized by the nearest-neighbor hopping amplitude $J>0$ and the on-site interaction energy $U>0$.
The quantities $\hat{a}_{R}$ and $\hat{a}^{\dagger}_{R}$ are, respectively, the annihilation and creation operators of a boson at the lattice site $R$, and $\hat{n}_{R}=\hat{a}^{\dagger}_{R}\hat{a}_{R}$ is the corresponding occupation number.
The average filling is $\bar{n}=\langle \hat{n}_R\rangle$ and we use unit lattice spacing ($R\in\mathbb{Z}$).
The equilibrium, zero-temperature phase diagram displays a Mott-insulating phase at integer fillings and sufficiently high values of $U/J$,
and a superfluid phase otherwise.
For unit filling in 1D, the critical interaction parameter is $U_\textrm{c}/J=3.3(1)$~\cite{kashurnikov1996exact,kuhner2000one,ejima2011dynamic,boeris2016mott}. 

We study the quench dynamics using a numerically exact time-dependent tensor network approach within time-dependent matrix product state ($t$-MPS) representation.
We typically use $L \simeq 96$ lattice sites and an evolution time of $t=10/J$, comparable with current experiments~\cite{trotzky2012probing,kohlert2019observation}.
The MPS bond and the local Hilbert-space dimensions, which are particularly demanding in the superfluid phase, are adjusted by checking the convergence of the numerical results. 

\begin{figure}[tb]
   \centering
    \includegraphics[scale=1]{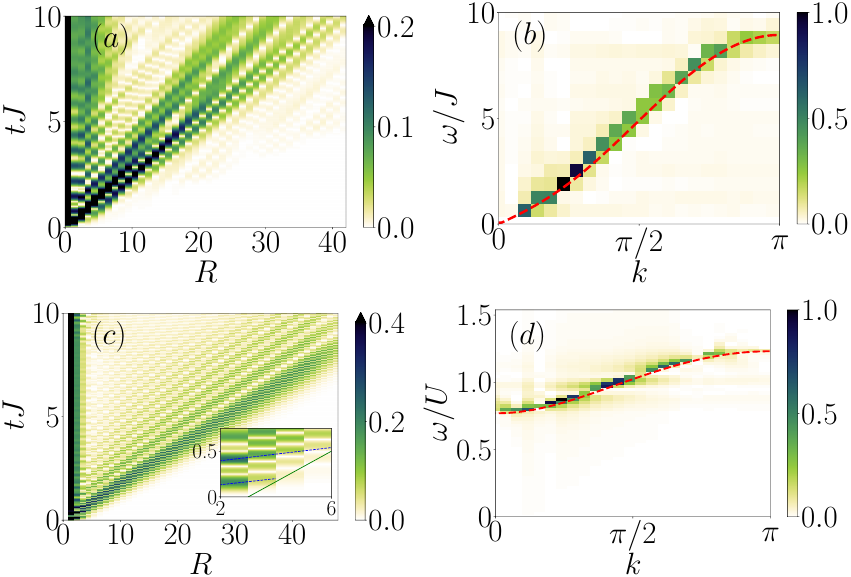}
   \caption{Benchmarking in the Bose-Hubbard chain.
   (a)~Absolute value of the two-body correlation function $G_2(R,t)$ obtained by $t$-MPS simulations for a global quench in the superfluid mean-field regime with $\bar{n}=5$, from $(U/J)_\textrm{i}=0.2$ to $U/J=0.1$ and (b)~Corresponding QSF, Eq.~\eqref{eq:QSF_spectral_form}, and comparison to the Bogoliubov dispersion relation, Eq.~\eqref{eq:DR_bogoliubov_SFMF} (dashed, red line).
   (c)~Same as (a) for $G_1(R,t)$ and a global quench in the strongly-interacting Mott phase with $\bar{n}=1$ from $(U/J)_\textrm{i}=25$ to $U/J=26$ (Inset: magnification).
   (d) QSF and comparison to the doublon-holon dispersion relation, Eq.~\eqref{eq:Mott_DR_doublon_holon} (dashed, red line).
   Note that the colorbars in (a) and (c) are cut off to improve visibility, the correlators being normalized by their maximum value.}
   \label{fig:benchmark}
\end{figure}

Figure~\ref{fig:benchmark}(a) shows the absolute value of the space-time evolution of the two-body correlation function $G_2(R,t)=\langle \delta\hat{n}(R,t)\delta\hat{n}(0,t)\rangle$ where $\delta\hat{n}(R,t)=\hat{n}(R,t)-\langle \hat{n}(R,t) \rangle$ for a quench at high filling, $\bar{n}=5$, from $(U/J)_\textrm{i}=0.2$ to $U/J=0.1$, both in the superfluid phase.
A characteristic linear cone-like propagation is clearly visible, outside which correlations decay exponentially~\cite{lieb1972finite}.
Inside the cone, the correlations show a complex space-time dependence.
Computing the space-time Fourier transform of $G_2(R,t)$, we find the QSF shown in Fig.~\ref{fig:benchmark}(b). As expected, it shows a sharp line, consistent with a well-defined dispersion relation of elementary excitations.
The result is in excellent quantitative agreement with the analytical prediction based on the Bogoliubov theory~\cite{pitaevskii2004},
\begin{equation}
\label{eq:DR_bogoliubov_SFMF}
\frac{2E_{k}}{J}=4\sqrt{2\sin^{2}(k/2)\left[2\sin^{2}(k/2)+\frac{\bar{n}U}{J}\right]},
\end{equation}
valid in the weakly-interacting regime, $\bar{n}\gg U/2J$.

The same analysis can be alternatively performed using the one-body correlation function $G_1(R,t)=\langle \hat{a}^{\dagger}(R,t)\hat{a}(0,t)\rangle$. While the result in real space and time is significantly blurred compared to the two-body correlation function, and the linear cone is hardly visible, the QSF allows us to extract the excitation spectrum with an accuracy comparable to Fig.~\ref{fig:benchmark}(b) (see Appendix~\ref{sec:suppl_G1_SFMF}).

We also performed the same analysis for a global quench in the strongly-interacting Mott phase at unit filling, $\bar{n}=1$, from $(U/J)_\textrm{i}=25$ to $U/J=26$.
The result for $G_1$ is shown in Figs.~\ref{fig:benchmark}(c) and (d).
The $G_1$ function again shows a linear cone, whose precise structure appears only on small time scales, see Inset of Fig.~\ref{fig:benchmark}(c).
The QSF, however, shows a sharp spectral branch, which compares very well with the doublon-holon pair dispersion relation~\cite{barmettler2012propagation},
\begin{equation}
\frac{2E_k}{U}\simeq\sqrt{\left[1-\frac{2J}{U} \! \left(2\bar{n}+1\right) \! \cos k\right]^{2}+\frac{16J^{2}}{U^{2}}\bar{n}(\bar{n}+1)\sin^{2}k}.
\label{eq:Mott_DR_doublon_holon}
\end{equation}
Note that, in contrast to the superfluid phase, choosing $G_1$ is instrumental for the Mott phase. This is because the ground state of the latter is nearly an eigenstate of the local density operator, $\hat{O}_{2}=\hat{n}$, and the couplings $\bra{m}\hat{O}_2\ket{0}$ in Eq.~\eqref{eq:QSF_spectral_form} are suppressed.
 
\section{Strongly-interacting superfluid regime}
\begin{figure}[t!]
   \centering
   \includegraphics[scale=1]{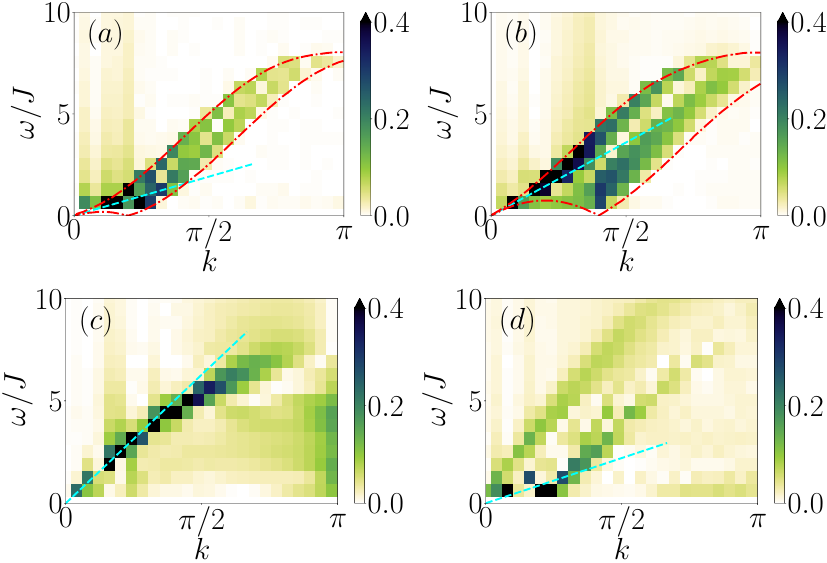}
   \caption{QSF of the Bose-Hubbard chain in the strongly-interacting superfluid regime.
   The quench is performed from $(U/J)_\textrm{i}=40$ to $U/J=50$ at (a)~$\bar{n}=0.1$, (b)~$\bar{n}=0.2$, (c)~$\bar{n}=0.5$, and (d)~$\bar{n}=0.9$. The dashed blue line represents a phonon branch propagating at $2v_\textrm{s}$.
   The approximate Bethe ansatz continuum (delimited by the dotted-dashed red lines) is shown in panels~(a) and (b).}
   \label{fig:SISF_QSF}
\end{figure}

Having validated the QSF approach to extract the excitation spectrum in the meanfield superfluid and Mott insulator limits, we now turn to the strongly-interacting superfluid regime, $U/J\bar{n} \gg 1$ and $\bar{n} \notin \mathbb{N}$, where no exact dispersion relation is known. The QSF probed by the $G_2$ correlation function for a quench to $U/J=50$ is shown in Fig.~\ref{fig:SISF_QSF} for increasing values of the filling factor $\bar{n}$.
It displays a broad but finite structure, which is easily interpreted within the continuous limit.
For low filling, $\bar{n}\ll1$, and long-wavelength excitations, $k\ll1$, the BHm may be mapped onto the Lieb-Liniger model, which is exactly solvable by Bethe ansatz~\cite{lieb1963exact1,lieb1963exact2}.
The excitation spectrum of the Lieb-Liniger model is a continuum delimited by two branches, called Lieb-I and Lieb-II modes, associated to particle-like and hole-like excitations, respectively.
We checked that for low filling [Figs.~\ref{fig:SISF_QSF}(a) and \ref{fig:SISF_QSF}(b)] the low $k$ sector of the QSF quantitatively agrees with the Lieb-Liniger spectrum (see Appendix~\ref{sec:suppl_LLBA_comparison}).
Yet the condition $k\ll1$ is very restrictive and the continuous Lieb-Liniger model is not sufficient to capture the breaking of convexity of the excitation branches observed in Fig.~\ref{fig:SISF_QSF}.

To overcome this issue, we developed an approximate Bethe ansatz (ABA) approach for the lattice model.
While the BHm is not exactly integrable for finite interactions, ABA approaches have been devised to compute the ground state properties of several models,
giving accurate results compared to exact numerical methods for low excitation densities~\cite{haldane1980solidification,krauth1991bethe,kiwata1994bethe}.
In the BHm, the breaking of integrability can be traced back to the presence of triply (or more) occupied sites~\cite{choy1982failure}.
For low filling, $\bar{n}\ll 1$, and strong interactions, $U/J\gg1$, the number of such highly occupied sites is strongly suppressed~\cite{note:exp_triple_occupancy} and we expect the ABA approach to be accurate.
This is consistent with Monte Carlo simulations comparing the complete and truncated BHm at zero temperature~\cite{kashurnikov1998zero}.

We compute the approximate excitation spectrum of the BHm extending the approach of Refs.~\cite{haldane1980solidification,choy1982failure,krauth1991bethe} and including particle-like and hole-like excitations,
similar to for the Lieb-Liniger model~\cite{lieb1963exact2}.
We force the many-body scattering to be factorized into two-body scattering processes.
The ABA yields a closed equation for the excitation backflow function, which is solved by an iterative algorithm.
The energy and the momentum of the two modes are then computed from this backflow (see Appendix~\ref{sec:suppl_ABA_BH}).
The possible excitations of the BHm combine a particle-like with a hole-like mode, which forms a continuum.
For a low filling $\bar{n}$, the boundaries of the latter, shown in red in Figs.~\ref{fig:SISF_QSF}(a) and \ref{fig:SISF_QSF}(b), and are in good agreement with the QSF results within the full Brillouin zone.

When $\bar{n}$ increases, many-body collisions become relevant and significantly alter the quasi-integrability of the model. The ABA approach breaks down and is not reported in Figs.~\ref{fig:SISF_QSF}(c) and (d).
Approaching half-filling, the two modes merge into a single, almost linear, branch, see Fig.~\ref{fig:SISF_QSF}(c). This branch is consistent with the phonon pair branch at the velocity $2v_\textrm{s}=4J$~\cite{cazalilla2004differences} (dashed blue line).
For higher fillings, a continuum is recovered within which two distinct, nearly linear excitation branches stand out, see Fig.~\ref{fig:SISF_QSF}(d).
Here however, they should not be confused with the phonon pair branch, which appears only at very low momentum, $k \ll 1-\bar{n}$, and has a significantly smaller velocity $2v_\textrm{s}\simeq 1.4 J$.
The upper linear branch corresponds to the fastest quasiparticles induced by the quench at the velocity $2v\simeq 4.8J$. It is consistent with the emergence of a unique characteristic velocity, faster than the speed of sound, in the vicinity of the causal cone as reported in Ref.~\cite{despres2019twofold} (see also Ref.~\cite{note:DoubleStructure}).

\section{Long-range interacting system}
\begin{figure}[t!]
   \centering
      \includegraphics[scale=1]{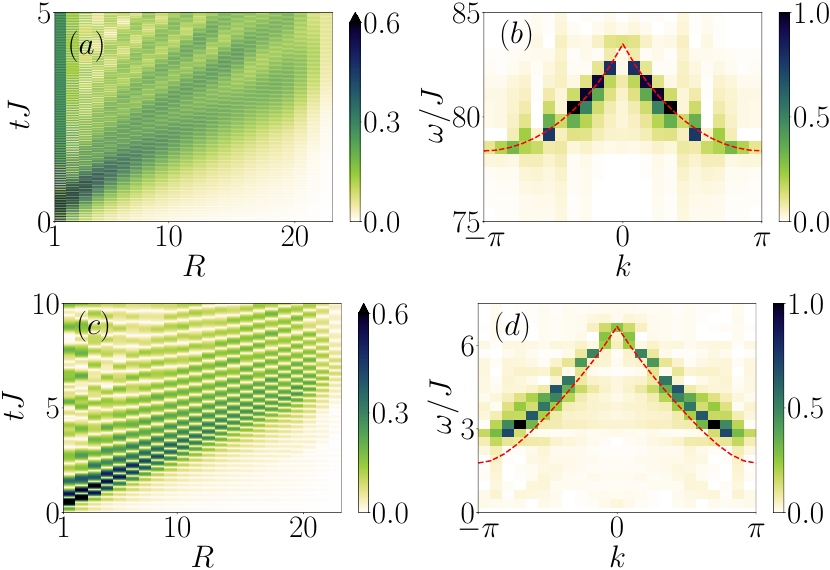}
   \caption{QSF for the LTRI chain.
(a)~Normalized absolute value of the spin correlation function $G^{xx}(R,t)$ obtained from $t$-MPS simulations for a quench at $\alpha=1.8$ from $(h/J)_{i}=50$ to $h/J=20$. (b)~Corresponding QSF and comparison with the LSWT prediction, Eq.~\eqref{eq:LSW_LRTI_disp_rel} (dotted red line). (c) and (d)~same as (a) and (b) respectively, for a quench to $h/J=1$.
}
   \label{fig:LRTI}
\end{figure}

Finally, we show that the QSF approach equally allows to probe the excitation spectrum of exotic models.
We illustrate this on the long-range transverse Ising (LRTI) chain, which can be realized experimentally using trapped ions~\cite{jurcevic2014,richerme2014} and has recently attracted significant attention~\cite{hauke2013spread,schachenmayer2013,cevolani2015protected,cevolani2016spreading,buyskikh2016,cevolani2018universal}.
The 1D Hamiltonian reads as
\begin{equation}
\hat{H}=\sum\limits_{R\neq R'}\dfrac{J}{\lvert R-R'\rvert ^{\alpha}} \hat{S}_{R}^{x}\hat{S}_{R'}^{x}-2h\sum\limits_{R}\hat{S}_{R}^{z},
\end{equation}
where $\hat{S}_{R}^{j}$ is the spin operator along the direction $j$ at site $R$,
$J$ is the spin exchange amplitude, and $h$ the magnetic field.
We perform quenches from $(h/J)_{i}=50$ to $h/J=20$ [Figs.~\ref{fig:LRTI}(a) and (b)] 
and $h/J=1$ [Figs.~\ref{fig:LRTI}(c) and (d)], 
and compute the spin correlation function $G^{xx}(R,t)=\langle \delta\hat{S}^{x}(R,t)\delta\hat{S}^{x}(0,t)\rangle$ with $\delta\hat{S}^{x}(R,t)=\hat{S}^{x}(R,t)-\langle \hat{S}^{x}(R,t) \rangle$ using $t$-MPS.
For both quenches, with $1<\alpha<2$, the spin correlations display a quasi-local cone, with algebraic leaks and a complex internal structure.
Instead, the QSF shows a sharp single-branch excitation spectrum. 
For the quench deep in the $z$-polarized phase, it is in excellent agreement with the linear spin-wave theory (LSWT) prediction~\cite{hauke2013spread,cevolani2016spreading},
\begin{equation}
\label{eq:LSW_LRTI_disp_rel}
\frac{2E_k}{J} = 4\sqrt{\frac{h}{J} \left[\frac{h}{J}+P_{\alpha}(k)\right]},
\end{equation}
with $P_{\alpha}(k)=\int \d R~\e^{-ikR}/\lvert R \rvert^{\alpha}$, see dashed red line in Fig.~\ref{fig:LRTI}(b).
For a stronger quench, to $h/J=1$, closer to the critical point at $(h/J)_\textrm{c}\sim 0.4$ \cite{koffel2012entanglement}, we still find a well-defined single excitation branch. It, however, shows significant deviations from the LSWT near the edges of the Brillouin zone, see Fig.~\ref{fig:LRTI}(d).

\section{Conclusion and outlook}
We have shown that the low-lying excitation spectrum of a many-body quantum system may be accurately extracted from the spectral representation of an ETC following a global quench via the QSF.
We explicitly demonstrated it for various 1D lattice models amenable to exact numerical calculations,
including Bose and spin models with short or long range interactions.
The approach is, however, general and applies equally well to  other systems,
\eg\ correlated fermions, continuous models, and in dimensions higher than one.

From an experimental point of view, the QSF approach may considerably simplify the measurement of excitation spectra in correlated systems compared to standard pump-probe spectroscopy techniques,
such as ARPES or Bragg spectroscopy.
The latter consists of exciting the system at well-defined frequency and wave vector, and observing the response of the system after some interaction time.
In practice, it requires to control the probe and systematically scan both the frequency and the wave vector.
In the QSF approach, the global quench replaces the pump. It generates
a complete set of excitations that propagate throughout the system,
by simply changing one parameter of the Hamiltonian.
At a given time $t$ after the quench, the spatial dependence of the ETC is measured by direct imaging of the full system, as now commonly done in AMO experiments.
For one-body correlators, this can be done by standard time-of-flight techniques.
For two-body correlators, it requires a series of images to measure density fluctuations.
Nevertheless, it avoids any tedious scan of probe parameters, in particular its momentum.
The full correlation pattern $G(\vec{R},t)$ is then obtained by scanning only $t$ from $0$ to some final time $T$.

Note that the QSF resolution is similar to that of standard approaches: the finite size of the system $L$ and the finite observation time $T$ used in experiments or numerical simulations typically lead to a spectral broadening of the QSF resonances of $\Delta k\sim 2\pi/L$ and $\Delta\omega\sim 2\pi/T$, respectively.
These effects can be straightforwardly included in the theory.
Moreover, the finite life time $\tau$ of the quasiparticles induces an additional frequency broadening $\Delta\omega\sim 2\pi/\tau$, which can be described by adding the Weisskopf-Wigner factor $\pm i/\tau$ to the quasiparticle energies in Eq.~(\ref{eq:QSF_spectral_form}).

\acknowledgments
This research was supported by the
European Commission FET-Proactive QUIC (H2020 Grant No.~641122).
The numerical calculations were performed using HPC resources 
from CPHT and GENCI-CCRT/CINES (Grant No.~c2017056853),
and making use of the ALPS library~\cite{dolfi2014matrix}.

 \appendix
 \section{Quench spectral function and dynamical structure factor}{\label{sec:suppl_QSFvsDSF}}

In the main paper, we consider a system in some nonstationary state represented by the density matrix $\hat{\rho}_\textrm{i}$, whose dynamics is governed by the Hamiltonian $\hat{H}$ at time $t>0$. We study the dynamics of the two-point correlator
\begin{equation}
\label{eq:SM_QSF0}
\begin{split}
G(\vec{x},\vec{y};t,t')& = \avg{\hat{O}_{1}^{\dagger}(\vec{x},t)\hat{O}_{2}(\vec{y},t')} \\
&=\text{Tr}\left[\hat{\rho}_{\textrm{i}}\hat{O}_{1}^{\dagger}(\vec{x},t)\hat{O}_{2}(\vec{y},t')\right],
\end{split}
\end{equation}
where $\hat{O}_{1}$ and $\hat{O}_{2}$ are local operators in the Heisenberg picture.
Note that here, beyond the precise scope of this paper, we consider the most general case of a possible unequal-time correlator ($t \neq t'$).
Since the dynamics of the operators in the Heisenberg picture is governed by the Hamiltonian $\hat{H}$,
it is convenient to use an eigenstate basis $\left\{\ket{n'}\right\}$ of the latter to compute the trace and insert two completeness relations, $\sum_{n} \vert n\rangle \langle n\vert=\sum_{m} \vert m\rangle \langle m\vert=1$, in Eq.~\eqref{eq:SM_QSF0}. Setting $\hbar=1$, we find
\begin{equation}
\label{eq:SM_QSF_space_dependence_not_extracted}
\begin{split}
G(\vec{x},\vec{y};t,t')&=\sum\limits_{n,n',m}\rho_{\mathrm{i}}^{n'n}\e^{i(E_{n}t-E_{n'}t')}\e^{-iE_{m}(t-t')}\\
&\quad\times\bra{n}\hat{O}_{1}^{\dagger}(\vec{x})\ket{m}\bra{m}\hat{O}_{2}(\vec{y})\ket{n'},
\end{split}
\end{equation}
where the operators $\hat{O}_{1,2}$ are now written in the Schr\"odinger picture and the time dependence disappears.

We now consider a translation invariant system. Using the translation operator from the origin to $\vec{x}$, we have $\hat{O}_j(\vec{x}) =\e^{-i\hat{\vec{P}} \cdot \vec{x}}\hat{O}_j(\vec{0})\e^{+i\hat{\vec{P}} \cdot \vec{x}}$,
where $\hat{\vec{P}}$ is the total momentum operator.
Moreover, we can use an eigenbasis common to $\hat{H}$ and $\hat{\vec{P}}$, so each eigenstate $\vert n\rangle$ has a well defined momentum $\vec{P}_{n}$.
Equation~(\ref{eq:SM_QSF_space_dependence_not_extracted}) then reads as
\begin{equation}
\label{eq:SM_QSF_not_simplified}
\begin{split}
G(\vec{x},\vec{y};t,t')&=\sum\limits_{n,n',m}\rho_{\mathrm{i}}^{n'n}\e^{i(E_{n}t-E_{n'}t')}\e^{-iE_{m}(t-t')}\\
&\quad\times\e^{i(\vec{P}_{m}-\vec{P}_{n})(\vec{x}-\vec{y})}\e^{i(\vec{P}_{n'}-\vec{P}_{n})\vec{y}}\\
&\quad\times\bra{n}\hat{O}_{1}^{\dagger}\ket{m}\bra{m}\hat{O}_{2}\ket{n'},\\
\end{split}
\end{equation}
where $\hat{O}_{j}$ is a short form for $\hat{O}_{j}(\vec{0},0)$.
Since the correlator $G$ only depends on $\vec{x}-\vec{y}$, it is convenient to
use the coordinates $\vec{R}=\vec{x}-\vec{y}$ and $\vec{r}=(\vec{x}+\vec{y})/2$, and write
$G(\vec{R};t,t') \equiv\frac{1}{L^D}\int\d \vec{r}\, G(\vec{r}+\vec{R}/2, \vec{r}-\vec{R}/2; t, t')$, where $L^D$ is the volume of the system in dimension $D$. Equation~\eqref{eq:SM_QSF_not_simplified} becomes
\begin{equation}
\label{eq:SM_QSF}
\begin{split}
G(\vec{R};t,t')
&=\left(\frac{2\pi}{L}\right)^{D} \sum\limits_{n,n',m}\delta(\vec{P}_{n}-\vec{P}_{n'})~\rho_{\mathrm{i}}^{n'n}\\
&\quad\times\e^{i(E_{n}t-E_{n'}t')}\e^{-iE_{m}(t-t')}\e^{i(\vec{P}_{m}-\vec{P}_{n})\vec{R}}\\
&\quad\times\bra{n}\hat{O}_{1}^{\dagger}\ket{m}\bra{m}\hat{O}_{2}\ket{n'}.
\end{split}
\end{equation}
Below, we separately examine the cases of the quench spectral function which is associated with an equal-time correlator $(t=t')$, and of the dynamical structure factor which is associated to an unequal-time correlator $(t\neq t')$, see Secs.~\ref{sec:SM_derivation_QSF} and \ref{sec:SM_derivation_DSF}, respectively.

\subsection{Quench spectral function}
\label{sec:SM_derivation_QSF}

\subsubsection{Derivation}
The QSF is defined as the space-time Fourier transform of an ETC and an out-of-equilibrium initial state. It corresponds to $\hat{\rho}_\textrm{i}$ such that $[\hat{\rho}_\textrm{i}, \hat{H}] \neq 0$ and $t=t'$ in Eq.~\eqref{eq:SM_QSF}. We then write
\begin{equation}
\label{eq:SM_QSF_simplified}
\begin{split}
G(\vec{k}&,\omega):=\int\d \vec{R} \, \d t\, \e^{-i\vec{k}\vec{R}-i\omega t} \, G(\vec{R};t,t)\\
&=\frac{(2\pi)^{2D+1}}{L^{D}}\sum\limits_{n,n',m}\rho_{\mathrm{i}}^{n'n}\bra{n}\hat{O}_{1}^{\dagger}\ket{m}\bra{m}\hat{O}_{2}\ket{n'}\\
&\quad\times\delta(E_{n}-E_{n'} - \omega)\delta(\vec{P}_{m}-\vec{P}_{n'}- \vec{k})\delta(\vec{P}_{n}-\vec{P}_{n'}),
\end{split}
\end{equation}
which is equivalent to Eq.~(\ref{eq:QSF_spectral_form}) of the main paper.

For a weak quench as considered in the main text, the initial state is close to the ground state $\ket{0}$,
so
\begin{equation}
\label{eq:SM_WeakQuench}
\hat{\rho}_{\mathrm{i}}\simeq\rho_{\mathrm{i}}^{00}\ket{0}\bra{0}+\sum\limits_{n\neq 0}\rho_{\mathrm{i}}^{0n}\ket{0}\bra{n}+\rho_{\mathrm{i}}^{n0}\ket{n}\bra{0}.
\end{equation}
For instance, a pure initial state close to the ground state is represented by $\ket{\psi_\textrm{i}}\simeq\ket{0}+\sum_{n\neq 0}\epsilon_{n}\ket{n}$ with $\epsilon_n \ll 1$,
and we find Eq.~(\ref{eq:SM_WeakQuench}) with $\rho_{\mathrm{i}}^{n0} = \left(\rho_{\mathrm{i}}^{0n}\right)^*=\epsilon_n$.
Therefore the only nonvanishing terms ${\rho}_{\mathrm{i}}^{n'n}$ correspond to either $n=0$ or $n'=0$ in Eq.~\eqref{eq:SM_QSF_simplified}, and the QSF simplifies into
\begin{widetext}
\begin{equation}
\label{eq:SM_QSF_weak_quench}
\begin{split}
G(\vec{k},\omega)&\simeq(2\pi)^{D+1}\sum\limits_{m}\rho_{\mathrm{i}}^{00}\bra{0}\hat{O}_{1}^{\dagger}\ket{m}\bra{m}\hat{O}_{2}\ket{0}\delta(\omega)\delta(\vec{P}_{m}-\vec{k})\\
&\qquad +\frac{(2\pi)^{2D+1}}{L^{D}}\sum\limits_{n,m}\rho_{\mathrm{i}}^{n0}\delta(\vec{P}_{n})\bra{0}\hat{O}_{1}^{\dagger}\ket{m}\bra{m}\hat{O}_{2}\ket{n}\delta(E_{n}+\omega)\delta(\vec{P}_{m}-\vec{k})\\
&\qquad +\frac{(2\pi)^{2D+1}}{L^{D}}\sum\limits_{n,m}\rho_{\mathrm{i}}^{0n}\delta(\vec{P}_{n})\bra{n}\hat{O}_{1}^{\dagger}\ket{m}\bra{m}\hat{O}_{2}\ket{0}\delta(E_{n}-\omega)\delta(\vec{P}_{m}-\vec{k}).\\
\end{split}
\end{equation}
\end{widetext}
Note that the momentum of the ground state is zero for symmetry reasons, $\vec{P}_{0}=\vec{0}$.
The first term in Eq.~(\ref{eq:SM_QSF_weak_quench}) is space and time independent and thus irrelevant for the dynamics. The last two terms include a resonance at negative and positive frequencies, respectively, associated to the Dirac distributions $\delta(E_{n} \pm \omega)$. In the main paper and in the following, we focus on the positive frequency sector were only the last term is relevant.

We now detail the selection rules mentioned in the main text, which allow to probe the excitation spectrum.
For weakly coupling operators, we can restrict the intermediate states $\ket{m}$ to single quasiparticles excitations (see Sec.~\ref{sec:examples_weak_operators}). The term $\delta(\vec{P}_m -\vec{k})$ imposes that $\ket{m}=\hat{b}^{\dagger}_{\vec{k}}\ket{0} \equiv \ket{\vec{k}}$, where $\hat{b}_{\vec{k}}^{\dagger}$ is the creation operator of a quasiparticle of momentum $\vec{k}$.
Owing to the term $\delta(\vec{P}_n)$, the first non-zero contribution is given by states $\ket{n}$ composed of two quasiparticles of opposite momenta, $\ket{n}=\hat{b}^{\dagger}_{-\vec{k}}\hat{b}^{\dagger}_{\vec{k}}\ket{0}$,
and energy $E_n=2E_\vec{k}$.
It finally yields
\begin{equation}
\label{eq:interpretation_QSF}
G(\vec{k},\omega>0) = \sum_{\vec{k}} \mathcal{F}(\vec{k})\delta(2E_{\vec{k}}-\omega),
\end{equation}
where the coefficient $\mathcal{F}(\vec{k})$ depends on the operators $\hat{O}_{1}$ and $\hat{O}_{2}$, and on the quench through the initial density matrix coefficients $\rho_{\mathrm{i}}^{0n}$.
Equation~\eqref{eq:interpretation_QSF} justifies the interpretation of the QSF as a direct probe of the excitation spectrum, through the resonance frequencies $\omega=2E_{\vec{k}}$.

\subsubsection{Weakly coupling operators}
\label{sec:examples_weak_operators}
In most cases of interest, the operators $\hat{O}_{1}$ and $\hat{O}_{2}$ can only create or annihilate a single quasiparticle excitation, and we refer to them as \textit{weakly coupling operators}.
This applies to a large number of situations, in particular all those considered in this paper, as detailed below.

Consider first the one-body correlation function,
\begin{equation}
\begin{split}
g_1(\vec{R},t)&=\avg{\hat{a}^{\dagger}(\vec{x}+\vec{R},t)\hat{a}(\vec{x},t)}\\
&=\sum_{\vec{k}}\e^{-i\vec{k}\cdot\vec{R}}\avg{\hat{a}_{\vec{k}}^{\dagger}(t)\hat{a}_{\vec{k}}(t)},
\end{split}
\end{equation}
where $\hat{a}_{\vec{k}}$ is the annihilation operator of a particle with momentum $\vec{k}$.
It corresponds to the correlation function $G(\vec{R},t)$ considered in the main paper [Eq.~\eqref{eq:observables}] with the single-particle operators $\hat{O}_1=\hat{O}_2=\hat{a}$.
The operator $\hat{O}_j$ may now be represented in terms of the single-quasiparticle operators.
A quasiparticle of momentum $\vec{k}$ representing a particle excitation dressed by other particles or holes
is associated to an annihilation operator $\hat{b}_\vec{k}$, which is a linear combination of the operators $\hat{a}_\vec{k}$ and $\hat{a}_\vec{k}^\dagger$. Reciprocally, the operators $\hat{a}_\vec{k}$ are linear combinations of the operators $\hat{b}_\vec{k}$ and $\hat{b}_\vec{k}^\dagger$.
Hence, the operator $\hat{O}_j$ can only create or annihilate a single quasiparticle. Therefore, the ground state $\ket{0}$ can only be coupled to a single-quasiparticle state, as assumed to derive Eq.~\eqref{eq:interpretation_QSF}.

For instance, the Bogoliubov quasiparticles representing the collective excitations of a Bose-Einstein condensate are related to the particle operators by
\begin{equation}
\label{eq:bogoliubov}
\begin{split}
\hat{a}_{\vec{k}}=u_\vec{k} \hat{b}_{\vec{k}}+v_\vec{k} \hat{b}_{-\vec{k}}^{\dagger},
\end{split}
\end{equation}
where $u_{\vec{k}}$ and $v_{\vec{k}}$ are the solutions of the Bogoliubov-de Gennes equations~\cite{pitaevskii2004}. 
A similar linear expression relating single-particle operators to single-quasiparticle operators also holds for doublon and holon excitations in the strongly interacting Mott phase of the Bose-Hubbard model, see for instance Ref.~\cite{barmettler2012propagation}.

More generally, higher-order operators can be cast in a similar form with generic hydrodynamic formulations.
Consider for instance the two-body correlation function
\begin{equation}
\begin{split}
g_2(\vec{R},t) &= \avg{\hat{n}(\vec{x}+\vec{R},t)\hat{n}(\vec{x},t)}\\
&= \sum_{\vec{k}}\e^{-i\vec{k}\cdot\vec{R}}\avg{\hat{n}_{\vec{k}}^{\dagger}(t)\hat{n}_{-\vec{k}}(t)}.
\end{split}
\end{equation}
It corresponds to the correlation function $G(\vec{r},t)$ considered in the main paper for the density operators $\hat{O}_1=\hat{O}_2=\hat{n}$.
The density operator may be expanded as $\hat{n}=n_{0}+\delta\hat{n}$ where $n_0$ is a classical field and $\delta\hat{n}$ represents the density fluctuations. The operator $\delta\hat{n}$ can be written, in momentum space,
\begin{equation}
\delta \hat{n}_{\vec{k}} = A_\vec{k} \left(  \hat{b}_{\vec{k}} + \hat{b}_{\vec{k}}^\dagger \right),
\end{equation}
see, for instance, Ref.~\cite{pitaevskii2004}. Similar to the one-body correlation function, the two-body correlation function can thus be decomposed in quasiparticle operators.
For instance, the hydrodynamic formulation may be used to describe a weakly-interacting Bose gas. Within Bogoliubov theory, one finds
\begin{equation}
\label{eq:justif_density_linear_operator}
A_\vec{k} = u_\vec{k} + v_\vec{k},
\end{equation}
where the quantities $u_{\vec{k}}$ and $v_{\vec{k}}$ are still the solutions of the Bogoliubov-De Gennes equations.
Note that this applies to both condensates~\cite{pitaevskii2004} and quasi-condensates~\cite{popov1983,mora2003}.
More generally, the hydrodynamic formulation may be applied to many correlated systems.
For instance, a similar form holds for 1D Luttinger liquids~\cite{giamarchi2004}.
Notice also that the phase operator, which is the conjugate of the density operator, can also be expanded in terms of single-quasiparticle operators.
  
Finally, for spin models in a polarized phase, for instance the LRTI model considered in this paper, the Holstein-Primakoff transformation can be used to map each spin operator onto bosonic operators. This transformation considers small deviations with respect to the mean-field ground state $(\avg{\hat{a}^{\dagger}(\vec{R})\hat{a}(\vec{R})}\ll 1$ for a spin $1/2$). It permits to map the spin operator in the direction orthogonal to the polarization axis into a single-particle bosonic one as~\cite{holstein1940,auerbach1994}
\begin{equation}
 \hat{S}^{x}_\vec{R}\simeq\dfrac{\hat{a}(\vec{R})+\hat{a}^{\dagger}(\vec{R})}{2}.
\end{equation} 
In terms of these bosonic variables, the Hamiltonian is quadratic and can therefore be diagonalized by introducing the linear Bogoliubov transformation in the form of Eq.~\eqref{eq:bogoliubov}.
Hence, for a spin correlation function as considered in the paper, the relevant operators are linear in the quasiparticle annihilation and creation operators.

\subsection{Comparison to the dynamical structure factor}
\label{sec:SM_derivation_DSF}
For the sake of comparison, we now consider dynamical structure factors (DSF), which are the quantities typically measured in pump-probe spectroscopy. The latter exploits the linear response induced by a weak perturbation of a system at equilibrium~\cite{mahan2000,bruus2004,pitaevskii2004}. The dynamical susceptibility (more precisely its imaginary part) is related to the DSF
\begin{equation}
\label{eq:SM_DSF_v1}
\begin{split}
G(\vec{k},\omega)&=2\pi\sum\limits_{n,m}\rho^{n n}\bra{n}\hat{O}_{1}^{\dagger}(\vec{k})\ket{m}\bra{m}\hat{O}_{2}(\vec{-k})\ket{n}\\
&\quad\times\delta(E_{n}-E_{m}+\omega)\\
&=(2\pi)^{D+1}\sum\limits_{n,m}\rho^{nn}\bra{n}\hat{O}_{1}^{\dagger}\ket{m}\bra{m}\hat{O}_{2}\ket{n}\\
&\quad\times\delta(\vec{P}_{m}-\vec{P}_{n}-\vec{k})\delta(E_{n}-E_{m}+\omega).\\
\end{split}
\end{equation}
The most usual case is that of equal operators, $\hat{O}_1=\hat{O}_2$, where the DSF reads as
\begin{equation}
\label{eq:SM_DSF_v2}
\begin{split}
G(\vec{k},\omega)=2\pi\sum\limits_{n,m}\rho^{n n} \lvert \bra{n}\hat{O}^{\dagger}(\vec{k})\ket{m} \rvert ^{2}\delta(E_{n}-E_{m}+\omega).\\
\end{split}
\end{equation}
Equation~(\ref{eq:SM_DSF_v1}) is nothing but the spectral representation (space and time Fourier transform)
\footnote{Here we use the usual convention $\int\d \vec{R}\d\tau \e^{-i\vec{k}\vec{R}+i\omega \tau}G(\vec{R};\tau)$. For the QSF we used another convention [see Eq.~\eqref{eq:SM_QSF_simplified}] which appears more convenient.}
of Eq.~(\ref{eq:SM_QSF_space_dependence_not_extracted}) for an UTC, that is, the spectral representation  of
\begin{equation}
\label{eq:SM_QSF_space_dependence_not_extracted_bis}
\begin{split}
G(\vec{x},\vec{y};t,t')=\sum\limits_{n,m}\rho^{nn} \bra{n}\hat{O}_{1}^{\dagger}(\vec{x},t)\ket{m}\bra{m}\hat{O}_{2}(\vec{y},t')\ket{n}.\\
\end{split}
\end{equation}
It is worth noting that linear response theory implies that the same, unperturbed Hamiltonian $\hat{H}$ governs both the initial state and the time evolution of the operators $\hat{O}_1$ and $\hat{O}_2$ in Eq.~(\ref{eq:SM_QSF_space_dependence_not_extracted_bis})~\cite{mahan2000,bruus2004,pitaevskii2004}.
In usual cases, the system is taken at thermodynamic equilibrium, where $\hat{\rho}=\exp(-\beta \hat{H})/Z$ with $Z=\textrm{Tr}[\exp(-\beta \hat{H})]$, in the canonical ensemble.
More generally, it is suffisant to assume that the state is stationary, \ie\ $[\hat{\rho}, \hat{H}]=0$.
In this case, the dynamical selection rule  $\delta(E_{n}-E_{m}+\omega)$ in Eqs.~(\ref{eq:SM_DSF_v1})-(\ref{eq:SM_DSF_v2}) is a direct consequence of the fact that the relevant correlator in real space and time representation, Eq.~(\ref{eq:SM_QSF_space_dependence_not_extracted_bis}), is an UTC, \ie\ with $t \neq t'$.
This is the main difference with the QSF discussed in Sec.~\ref{sec:SM_derivation_QSF} where the state is out of equilibrium and the dynamical selection rule emerges from the spectral representation of an ETC.

It is also worth noting that while both the quench spectral function and the dynamical structure factor allow us to determine the quasiparticle dispersion relation through dynamical selection rules, they are different quantities.
To illustrate this, consider the DSF, assuming for simplicity that the system is in the ground state, so Eq.~(\ref{eq:SM_DSF_v1}) reads as
\begin{equation}
\label{eq:SM_DSF_zeroT}
\begin{split}
G(\vec{k},\omega)&=(2\pi)^{D+1}\sum\limits_{m}\rho^{0 0}\bra{0}\hat{O}_{1}^{\dagger}\ket{m}\bra{m}\hat{O}_{2}\ket{0}\\
&\quad\times\delta(\vec{P}_{m}-\vec{k})\delta(\omega-E_{m}).
\end{split}
\end{equation}
Then, as for the QSF, the quasiparticle dispersion relation appears assuming that  $\hat{O}_1$ and $\hat{O}_2$ are weakly coupling operators \ie~they couple the ground state $\ket{0}$ only to single-quasiparticle states of the form $\ket{m}=\ket{\vec{k}}$ of momentum $\vec{k}$ and energy $E_m=E_\vec{k}$.
In this case, the DSF peaks at $\omega=E_\vec{k}$, hence providing the quasiparticle dispersion relation. In contrast, the QSF couples the ground state to a single-quasiparticle states $\ket{m}=\ket{\vec{k}}$ and then to a pair of quasiparticles with opposite momenta and same energies, so that the QSF peaks at $\omega=2E_\vec{k}$, see Sec.~\ref{sec:SM_derivation_QSF}.
 
 \section{Quench spectral function for the $\boldsymbol{G_1}$ correlation function in the superfluid mean-field regime}
\begin{figure}[h!]
   \centering
   \includegraphics[width=0.47\textwidth]{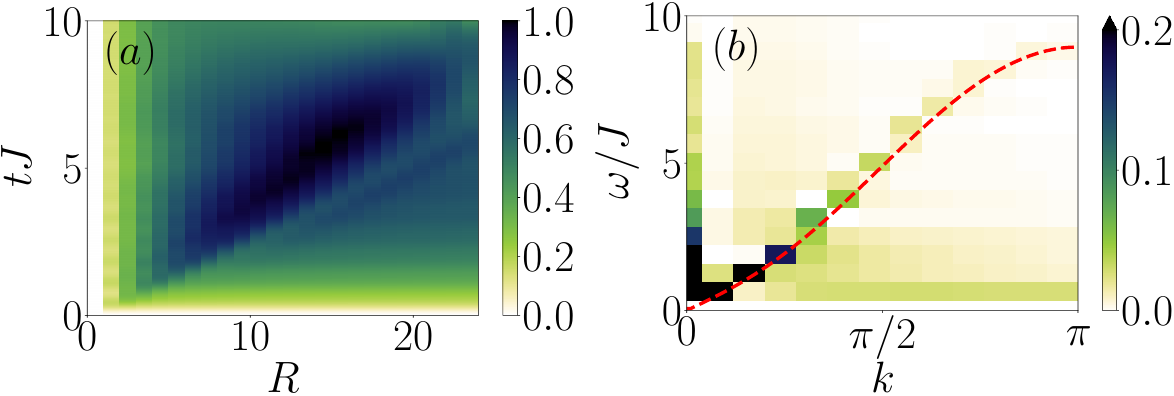}
   \caption{(a) Absolute value of the space-time evolution of $G_1(R,t)$ obtained by t-MPS calculations after a global quench in the superfluid mean-field regime with $\bar{n}=5$, from $U/J=0.2$ to $0.1$.
   (b) Associated QSF and comparison to the Bogoliubov dispersion relation, Eq.~\eqref{eq:DR_bogoliubov_SFMF} of the main text (dashed red line). Note that the colorbar in (b) is cut off to 20\% to improve visibility, and the correlator is normalized by its maximum value.}   
  \label{fig:supplemental_MFSF_G1}
\end{figure}

In the main text, we discussed the determination of the excitation spectrum in the superfluid mean-field regime from the QSF associated to the two-body correlation function $G_{2}(R,t)$. Here, we show the counterpart of this analysis for the one-body correlation function $G_{1}(R,t)$, computed using the same numerical approach and the same quench.
The $G_{1}(R,t)$ function and the associated QSF are shown in Figs.~\ref{fig:supplemental_MFSF_G1}(a) and \ref{fig:supplemental_MFSF_G1}(b), respectively.
As observed in Fig.~\ref{fig:supplemental_MFSF_G1}(a), the $G_{1}(R,t)$ function is quite blurred owing to quasi-long-range correlations already present in the initial state. In particular, the causal cone is hardly visible here.
The associated QSF, however, displays a clear single branch, see Fig.~\ref{fig:supplemental_MFSF_G1}(b).
The latter is in good agreement with the Bogoliubov dispersion relation given by the Eq.~\eqref{eq:DR_bogoliubov_SFMF} of the main text (dashed red line).
It shows that the excitation spectrum can also be extracted from the one-body correlator in spite of a signal in real space and time that is significantly less sharp than for the two-body correlator.
 \label{sec:suppl_G1_SFMF}

\section{Comparison between the quench spectral function of the Bose-Hubbard model and the Lieb-Liniger modes in the continuous limit}
\label{sec:suppl_LLBA_comparison}
In the continuous limit, $\bar{n}\ll 1$, and for low-momentum excitations, $k \ll 1$, the Bose-Hubbard model can be mapped onto the Lieb-Liniger model,
\begin{equation}
\mc{H}=-\dfrac{\hbar^{2}}{2m}\sum_{i}\dfrac{\partial^{2}}{\partial x_{i}^{2}}+\dfrac{\hbar^{2}c}{m}\sum\limits_{i<j}\delta(x_i-x_j),
\end{equation}
where $m$ is the particle mass, $x_i$ is the position of the $i$-th particle, and $c$ (homogeneous to the inverse of a length) stands for the interaction strength.
The mapping is found by discretizing the wave function on a length scale $a$, associated to the lattice spacing of the Bose-Hubbard model.
It yields $U\equiv\hbar^{2}c/ma$, $J\equiv\hbar^{2}/2ma^{2}$, and thus $c\equiv U/2Ja$. 

\begin{figure}[tb!]
   \centering
   \includegraphics[width=0.47\textwidth]{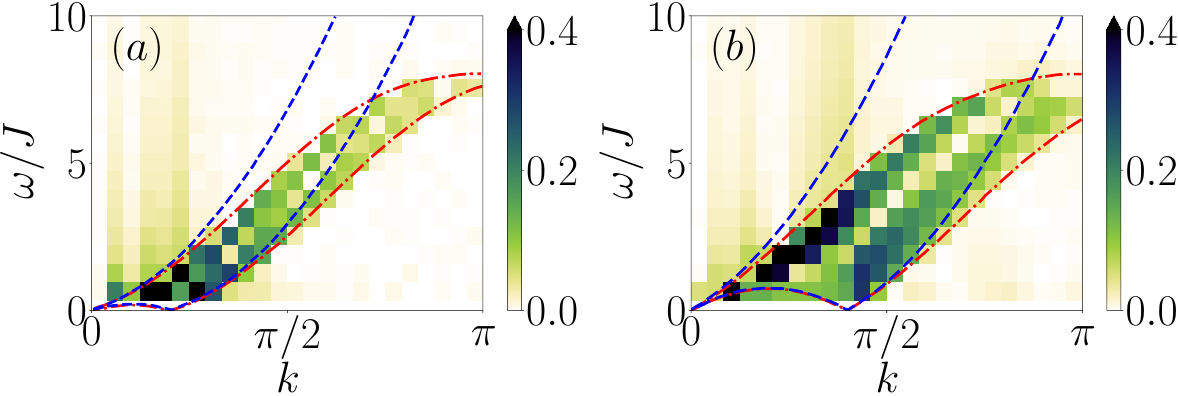}
   \caption{QSF (green) associated to the two-body correlation function for the Bose-Hubbard chain in the strongly-interacting superfluid regime, together with the Lieb branches of the continuous Lieb-Liniger model (dashed blue lines).
   The Lieb-like branches found using the approximate Bethe ansatz approach for the Bose-Hubbard model are also shown (dotted-dashed red lines).
The quench is performed from $(U/J)_\textrm{i}=40$ to $U/J=50$ at (a)~$\bar{n}=0.1$ and (b)~$\bar{n}=0.2$.}
   \label{fig:supplemental_SISF_QSF_LLBA_BHBA}
\end{figure}

The Lieb-Liniger model is known to be integrable by Bethe ansatz~\cite{lieb1963exact1,lieb1963exact2}.
Its excitation spectrum is a continuum delimited by the so-called Lieb-I (particle-like) and Lieb-II (hole-like) branches.
In Fig.~\ref{fig:supplemental_SISF_QSF_LLBA_BHBA}, we reproduce the Figs.~\ref{fig:SISF_QSF}(a) and \ref{fig:SISF_QSF}(b) of the main paper, showing the QSF for the Bose-Hubbard chain in the strongly interacting superfluid regime at low fillings,
together with the Lieb branches of the continuous Lieb-Liniger model (dashed blue lines).
For small momenta, $k\ll 1$, the two Lieb branches are in quantitative agreement with the QSF result (green) as well as with the predictions of the approximate Bethe ansatz (dotted-dashed red line, see below).
In contrast, for larger momenta, $k \gtrsim 1$, the lattice discretization becomes relevant and the Lieb branches deviate from the QSF result.

\section{Approximate Bethe ansatz for excitations in the Bose-Hubbard chain}
\label{sec:suppl_ABA_BH}
Here we outline the main steps for the derivation of the approximate Bethe ansatz (ABA) approach used in the main text. 
For a comprehensive introduction to the general Bethe ansatz formalism, see for instance Refs.~\cite{korepin1997quantum,sutherland2004beautiful,franchini2017introduction} and references therein.
The approach detailed below was originally developed in Ref.~\cite{krauth1991bethe} to derive the ground state properties of the BHm. We extend it to the derivation of the excitation spectrum.

We first review the ABA approach for the ground state, starting with two particles.
In one dimension, the particles can be ordered such that $x_1<x_2$ where $x_{j}$ is the position of the particle $j$. For bosons as considered here, the global wave function is symmetric under the exchange of coordinates,
and we may restrict the discussion to $x_1<x_2$ without loss of generality.
We take  $A(k_1,k_2)\e^{i(k_1x_1+k_2x_2)}+A(k_2,k_1)\e^{i(k_2x_1+k_1x_2)}$ as an ansatz for the reduced two-body wave function, with $k_j$ the quasimomentum of the particle $j$. The amplitudes $A(k_1,k_2)$ and $A(k_2,k_1)$ are the unknown coefficients, which we want to determine. The reduced wave function with $x_2<x_1$ is given by the same formula simply exchanging $x_1$ and $x_2$, keeping the same amplitudes.
As for any formulation of the Bethe ansatz, we impose that the energy involved in the time-independent Schr\"odinger equation is the one associated to free particles, and include the interaction in the way the quasimomenta are distributed. Here we use $E=-2J(\cos k_{1}+\cos k_2)$ as suggested by the Bose-Hubbard Hamiltonian with $U=0$. The ansatz for the reduced wave functions, the continuity of the global wave function at $x_1=x_2$ and the previous form of the energy can be simultaneously imposed if the following condition is satisfied:
\begin{equation}
\label{eq:SM_ABA_Smatrix}
\begin{split}
\dfrac{A(k_1,k_2)}{A(k_2,k_1)}=\dfrac{i(\sin k_1 -\sin k_2)-\frac{U}{2J}}{i(\sin k_1 -\sin k_2)+\frac{U}{2J}}:=-\e^{i\theta_{12}}.
\end{split}
\end{equation}
This result is found by working along the lines of Refs.~\cite{lieb1963exact1,lieb1963exact2}, adding the presence of the lattice in the formulation. This equation defines the scattering phase $\theta_{12}$ which can be rewritten conveniently as 

\begin{equation}
\label{eq:SM_ABA_Smatrix}
\begin{split}
\theta_{12}=-2\arctan\left[\frac{2J}{U}(\sin k_1 -\sin k_2)\right].
\end{split}
\end{equation}
This fully solves the problem in the case $N=2$.

We now turn to $N>2$ particles. The Bose-Hubbard model is not integrable for a finite interaction parameter $U/J$. This means that a many-body scattering cannot be factorized in an exact way as a product of two-body collisions.  Already including a third boson into the description cannot be done in an exact way \ie\ following the same procedure, as was first pointed out in Ref.~\cite{choy1982failure}. To be more specific, when the ansatz of the reduced wave function $\sum_{\mathcal{P}\in \mathcal{S}_{N}}A_{\mathcal{P}}\e^{ik_{\mathcal{P}_j}x_j}$, where $\mathcal{S}_N$ is the permutation group of $N!$ elements, and the form of the energy $E=-2J\sum_{i=1}^{N} \cos k_{i}$ are simultaneously imposed, the continuity of the global wave function when all $x_j$ are equal cannot be satisfied. Such pathological cases occur when at least three bosons interact in the same lattice site. For low densities and high interactions, however, such multi-occupied states are very strongly attenuated~\cite{ronzheimer2013expansion}.
In this regime, we expect that the previous description, now called approximate Bethe ansatz, yields a reasonable description. This point was originally pointed out in Ref.~\cite{krauth1991bethe}.
To adapt the $N=2$ solution to the thermodynamic limit, we impose periodic boundary conditions on the global wave function: $\forall j, \psi(x_1,...,x_j+L,...x_N)=\psi(x_1,...,x_j+L,...x_N)$.
Generalizing Eq.~\eqref{eq:SM_ABA_Smatrix}, this condition reads as
\begin{equation}
\e^{ik_j L}=\prod_{l\neq j}\dfrac{i(\sin k_j - \sin k_l) -\frac{U}{2J}}{i(\sin k_j - \sin k_l)+\frac{U}{2J}}=(-1)^{N-1}\e^{i\sum\limits_{l\neq j}\theta_{jl}},
\end{equation}
with $\theta_{jl}$ defined above. In log form, this gives the so-called Bethe equations for the BHm,
\begin{equation}
\label{eq:ABA_bethe_equation}
k_{j}L=2\pi I_{j}+\sum\limits_{l\neq j}\theta_{jl},
\end{equation}
 where $I_{j}$ are integers (for $N$ odd) evenly distributed between -$(N-1)/2$ and $+(N-1)/2$. This equation relates the quasimomentum distribution to the interactions through the scattering phase.
Moving to the thermodynamic limit, we introduce the quasimomentum density
 \begin{equation}
\rho(k_j)=\lim_{N,L\rightarrow\infty}\dfrac{1}{L(k_{j+1}-k_j)}.
 \end{equation}
 We then take the difference between Eq.~\eqref{eq:ABA_bethe_equation} for $k_{j+1}$ and $k_{j}$, respectively. Considering $k$ as a continuous variable, one then shows that the quasimomentum density $\rho(k)$ obeys the linear integral equation
\begin{equation}
2\pi\rho(k)=1+\dfrac{U}{J}\cos k \int_{-k_\textrm{F}}^{+k_\textrm{F}}\dfrac{\rho(k')\d k'}{\left(\frac{U}{2J}\right)^{2}+\left(\sin k-\sin k'\right)^{2}},
\label{eq:BHBA_bethe_eq1}
\end{equation}
where the Fermi momentum $k_\textrm{F}$ is determined by the density $n$ through the relation
\begin{equation}
\label{eq:BHBA_bethe_eq2}
n=\int_{-k_\textrm{F}}^{+k_\textrm{F}}\rho(k)~\d k.
\end{equation}
From these two equations, all ground state quantities can be computed. 
In practice, we fix the density $n$ and the interaction parameter $U/2J$,
and solve iteratively Eqs.~(\ref{eq:BHBA_bethe_eq1}) and (\ref{eq:BHBA_bethe_eq2})
for both $k_\textrm{F}$ and $\rho(k)$ until convergence has been reached. 
The equations \eqref{eq:BHBA_bethe_eq1} and \eqref{eq:BHBA_bethe_eq2} were first derived in Ref.~\cite{haldane1980solidification} (see also Ref.~\cite{krauth1991bethe} for the correction of a typo about a factor of 2). 

We now extend the ABA approach to the determination of the low-energy number-conserving excitations. They are found by removing one quasimomentum $k_\textrm{h}$ from (\ie\ create a hole in) the Fermi sea, such that $-{k_\textrm{F}}\leq k_\textrm{h} \leq {k_\textrm{F}}$, and put it back (create a particle) above the Fermi level at $k_\textrm{p}$, such that $-\pi <  k_\textrm{p} <-{k_\textrm{F}}$ or ${k_\textrm{F}} <  k_\textrm{p} \leq \pi$.
This excited state is characterized by the new values of the quasimomenta $k_j^{\star}$, which are distributed according to the Bethe Eq.~\eqref{eq:ABA_bethe_equation} for $k_{j}^{\star}$.  
In analogy with Refs.~\cite{lieb1963exact2,yang1969thermodynamics}, we introduce a backflow function. It accounts, to first order, for the redistribution of the quasimomenta between the excited state and the ground state. Its expression is $\mc{J}(k_{j})=L\rho(k_{j})\Delta k_{j}$ where $\Delta k_{j}=k_{j}^{\star}-k_{j}$ is assumed to be small compared to $k_j$. Taking the difference between the Bethe equations for $k_{j}^{\star}$ and $k_{j}$ yields, in the thermodynamic limit, the following linear equation for the backflow function:
\begin{equation}
\begin{split}
\label{eq:ABA_backflow}
2 \pi \mc{J}(k)&=\theta(\sin k - \sin  k_\textrm{p})-\theta(\sin k - \sin k_\textrm{h})
\\
&\quad+\dfrac{U}{J}\int_{-{k_\textrm{F}}}^{+{k_\textrm{F}}}\dfrac{\mc{J}(k')\cos k' \d k'}{\left(\dfrac{U}{2J}\right)^{2}+\left[\sin k - \sin k' \right]^{2}}.
\end{split}
\end{equation}
Equation~\eqref{eq:ABA_backflow} can be solved numerically for a given excitation $(k_\textrm{h},k_\textrm{p})$ and with ${k_\textrm{F}}$ determined previously for a given set of the physical parameters $n$ and $U/J$. We can then compute the momentum difference $\Delta P$ between the excited state and the ground state, and their energy difference $\Delta E$ in terms of this backflow:
\begin{equation}
\label{eq:ABA_excitation_spectrum}
 \begin{split}
 \Delta P&=\sum_{j}(k_{j}^{\star}-k_j)\overset{\text{TL}}{=} k_\textrm{p}-k_\textrm{h}+\int_{-{k_\textrm{F}}}^{+{k_\textrm{F}}}\mc{J}(k)\d k, \\
 \dfrac{\Delta E}{J}&=-2\sum_{j}(\cos k_{j}^{\star}-\cos k_{j})\\
 &\overset{\text{TL}}{=} -2\cos k_\textrm{p}+2\cos k_\textrm{h}+2\int_{-{k_\textrm{F}}}^{+{k_\textrm{F}}}\mc{J}(k)\sin k~\d k.
 \end{split}
\end{equation} 
where $\overset{\text{TL}}{=}$ stands for the thermodynamic limit.
The continuum 
of excitations $(\Delta P,\Delta E)$ is computed numerically by first solving Eq.~\eqref{eq:ABA_backflow}  and then varying $-k_\textrm{F}\leq k_\textrm{h}\leq {k_\textrm{F}}$, and $-\pi\leq k_\textrm{p} < -k_\textrm{F}$ or $k_\textrm{F}< k_\textrm{p} \leq \pi$.

Note that, in the limit $U/J\rightarrow\infty$, we find $\theta\rightarrow 0$ and $J(k)\rightarrow 0$ in Eq.~\eqref{eq:ABA_backflow}. We recover the well-known fully fermionized regime.
Moreover, in the continuous limit where the lattice spacing is set to 0, and therefore the quasimomenta $k\rightarrow 0$ (recall the quasimomentum is measured in units of the lattice spacing), we recover the known Bethe equations for the Lieb-Liniger model, see for instance Eqs.~(43), (45), and (46) in Ref.~\cite{yang1969thermodynamics} (written there for finite temperature). 



\bibliographystyle{apsrev4-1} 

\begin{thebibliography}{10}
\providecommand*{\bibinfo}[2]{#2}
\providecommand*{\eprint}[1]{#1}
\providecommand*{\url}[1]{#1}
\bibitem{negele1988}
\bibinfo{author}{J.~W. Negele} and \bibinfo{author}{H.~Orland},
  \bibinfo{title}{\emph{Quantum Many-Particle Systems}}
  (\bibinfo{publisher}{CRC Press, London}, \bibinfo{year}{1988}).
\bibitem{mahan2000}
\bibinfo{author}{G.~Mahan}, \bibinfo{title}{\emph{Many Particle Physics}}
  (\bibinfo{publisher}{Springer, New York}, \bibinfo{year}{2000}).
\bibitem{giamarchi2004}
\bibinfo{author}{T.~Giamarchi}, \bibinfo{title}{\emph{{Quantum Physics in One
  Dimension}}} (\bibinfo{publisher}{Carendon Press, {O}xford},
  \bibinfo{year}{2004}).
\bibitem{bruus2004}
\bibinfo{author}{H.~Bruus} and \bibinfo{author}{K.~Flensberg},
  \bibinfo{title}{\emph{{M}any-Body Quantum Theory in Condensed Matter Physics:
  {A}n Introduction}} (\bibinfo{publisher}{Oxford University Press, Oxford},
  \bibinfo{year}{2004}).
\bibitem{rickayzen2013}
\bibinfo{author}{G.~Rickayzen}, \bibinfo{title}{\emph{Green's Functions and
  Condensed Matter}} (\bibinfo{publisher}{Dover, New York},
  \bibinfo{year}{2013}).
\bibitem{caux2006dynamical}
\bibinfo{author}{J.-S. Caux} and \bibinfo{author}{P.~Calabrese},
  \bibinfo{title}{\emph{Dynamical density-density correlations in the
  one-dimensional {B}ose gas}}, \bibinfo{journal}{Phys. Rev. A}
  \bibinfo{volume}{\textbf{74}}(3), \bibinfo{pages}{031605(R)}
  (\bibinfo{date}{2006}).
\bibitem{pippan2009excitation}
\bibinfo{author}{P.~Pippan}, \bibinfo{author}{H.~G. Evertz}, and
  \bibinfo{author}{M.~Hohenadler}, \bibinfo{title}{\emph{Excitation spectra of
  strongly correlated lattice bosons and polaritons}}, \bibinfo{journal}{Phys.
  Rev. A} \bibinfo{volume}{\textbf{80}}(3), \bibinfo{pages}{033612}
  (\bibinfo{date}{2009}).
\bibitem{ejima2012dynamic}
\bibinfo{author}{S.~Ejima}, \bibinfo{author}{H.~Fehske}, and
  \bibinfo{author}{F.~Gebhard}, \bibinfo{title}{\emph{Dynamic density-density
  correlations in interacting {B}ose gases on optical lattices}},
  \bibinfo{journal}{J. Phys. Conf. Ser.} \bibinfo{volume}{\textbf{391}}(1),
  \bibinfo{pages}{012143} (\bibinfo{date}{2012}).
\bibitem{furrer2009}
\bibinfo{author}{A.~Furrer}, \bibinfo{author}{J.~Mesot}, and
  \bibinfo{author}{T.~Str\"ssle}, \bibinfo{title}{\emph{Neutron Scattering in
  Condensed Matter Physics}} (\bibinfo{publisher}{World Scientific, Singapore},
  \bibinfo{year}{2009}).
\bibitem{damascelli2004}
\bibinfo{author}{A.~Damascelli}, \bibinfo{title}{\emph{Probing the electronic
  structure of complex systems by {ARPES}}},
  \bibinfo{journal}{\Jphysicascripta} \bibinfo{volume}{\textbf{T109}},
  \bibinfo{pages}{61} (\bibinfo{date}{2004}).
\bibitem{stewart2008using}
\bibinfo{author}{J.~Stewart}, \bibinfo{author}{J.~Gaebler}, and
  \bibinfo{author}{D.~Jin}, \bibinfo{title}{\emph{Using photoemission
  spectroscopy to probe a strongly interacting {F}ermi gas}},
  \bibinfo{journal}{Nature} \bibinfo{volume}{\textbf{454}}(7205),
  \bibinfo{pages}{744} (\bibinfo{date}{2008}).
\bibitem{stenger1999bragg}
\bibinfo{author}{J.~Stenger}, \bibinfo{author}{S.~Inouye},
  \bibinfo{author}{A.~P. Chikkatur}, \bibinfo{author}{D.~M. Stamper-Kurn},
  \bibinfo{author}{D.~E. Pritchard}, and \bibinfo{author}{W.~Ketterle},
  \bibinfo{title}{\emph{Bragg spectroscopy of a {B}ose-{E}instein condensate}},
  \bibinfo{journal}{Phys. Rev. Lett.} \bibinfo{volume}{\textbf{82}}(23),
  \bibinfo{pages}{4569} (\bibinfo{date}{1999}).
\bibitem{ozeri2005colloquium}
\bibinfo{author}{R.~Ozeri}, \bibinfo{author}{N.~Katz},
  \bibinfo{author}{J.~Steinhauer}, and \bibinfo{author}{N.~Davidson},
  \bibinfo{title}{\emph{Colloquium: Bulk {B}ogoliubov excitations in a
  {B}ose-{E}instein condensate}}, \bibinfo{journal}{Rev. Mod. Phys.}
  \bibinfo{volume}{\textbf{77}}(1), \bibinfo{pages}{187}
  (\bibinfo{date}{2005}).
\bibitem{clement2009exploring}
\bibinfo{author}{D.~Cl{\'e}ment}, \bibinfo{author}{N.~Fabbri},
  \bibinfo{author}{L.~Fallani}, \bibinfo{author}{C.~Fort}, and
  \bibinfo{author}{M.~Inguscio}, \bibinfo{title}{\emph{Exploring correlated
  {1D} {B}ose gases from the superfluid to the {M}ott-insulator state by
  inelastic light scattering}}, \bibinfo{journal}{Phys. Rev. Lett.}
  \bibinfo{volume}{\textbf{102}}(15), \bibinfo{pages}{155301}
  (\bibinfo{date}{2009}).
\bibitem{meinert2015probing}
\bibinfo{author}{F.~Meinert}, \bibinfo{author}{M.~Panfil},
  \bibinfo{author}{M.~J. Mark}, \bibinfo{author}{K.~Lauber},
  \bibinfo{author}{J.-S. Caux}, and \bibinfo{author}{H.-C. N{\"a}gerl},
  \bibinfo{title}{\emph{Probing the excitations of a {L}ieb-{L}iniger gas from
  weak to strong coupling}}, \bibinfo{journal}{Phys. Rev. Lett.}
  \bibinfo{volume}{\textbf{115}}(8), \bibinfo{pages}{085301}
  (\bibinfo{date}{2015}).
\bibitem{polkovnikov2011}
\bibinfo{author}{A.~Polkovnikov}, \bibinfo{author}{K.~Sengupta},
  \bibinfo{author}{A.~Silva}, and \bibinfo{author}{M.~Vengalattore},
  \bibinfo{title}{\emph{\textit{Colloquium}: Nonequilibrium dynamics of closed
  interacting quantum systems}}, \bibinfo{journal}{\Jrmp}
  \bibinfo{volume}{\textbf{83}}, \bibinfo{pages}{863} (\bibinfo{date}{2011}).
\bibitem{eisert2015}
\bibinfo{author}{J.~Eisert}, \bibinfo{author}{M.~Friesdorf}, and
  \bibinfo{author}{C.~Gogolin}, \bibinfo{title}{\emph{Quantum many-body systems
  out of equilibrium}}, \bibinfo{journal}{\Jnatphys}
  \bibinfo{volume}{\textbf{11}}, \bibinfo{pages}{124} (\bibinfo{date}{2015}).
\bibitem{langen2015}
\bibinfo{author}{T.~{Langen}}, \bibinfo{author}{R.~{Geiger}}, and
  \bibinfo{author}{J.~{Schmiedmayer}}, \bibinfo{title}{\emph{Ultracold atoms
  out of equilibrium}}, \bibinfo{journal}{\JAnnualRevCondMat}
  \bibinfo{volume}{\textbf{6}}, \bibinfo{pages}{201} (\bibinfo{date}{2015}).
\bibitem{lewenstein2007}
\bibinfo{author}{M.~Lewenstein}, \bibinfo{author}{A.~Sanpera},
  \bibinfo{author}{V.~Ahufinger}, \bibinfo{author}{B.~Damski},
  \bibinfo{author}{A.~Sen}, and \bibinfo{author}{U.~Sen},
  \bibinfo{title}{\emph{Ultracold atomic gases in optical lattices: {M}imicking
  condensed matter physics and beyond}}, \bibinfo{journal}{\Jadvphys}
  \bibinfo{volume}{\textbf{56}}, \bibinfo{pages}{243} (\bibinfo{date}{2007}).
\bibitem{bloch2008}
\bibinfo{author}{I.~Bloch}, \bibinfo{author}{J.~Dalibard}, and
  \bibinfo{author}{W.~Zwerger}, \bibinfo{title}{\emph{Many-body physics with
  ultracold gases}}, \bibinfo{journal}{\Jrmp} \bibinfo{volume}{\textbf{80}},
  \bibinfo{pages}{885} (\bibinfo{date}{2008}).
\bibitem{NaturePhysicsInsight2012cirac}
\bibinfo{author}{J.~I. Cirac} and \bibinfo{author}{P.~Zoller},
  \bibinfo{title}{\emph{Goals and opportunities in quantum simulation}},
  \bibinfo{journal}{\Jnatphys} \bibinfo{volume}{\textbf{8}},
  \bibinfo{pages}{264} (\bibinfo{date}{2012}).
\bibitem{NaturePhysicsInsight2012bloch}
\bibinfo{author}{I.~Bloch}, \bibinfo{author}{J.~Dalibard}, and
  \bibinfo{author}{S.~Nascimb\`ene}, \bibinfo{title}{\emph{Quantum simulations
  with ultracold quantum gases}}, \bibinfo{journal}{\Jnatphys}
  \bibinfo{volume}{\textbf{8}}, \bibinfo{pages}{267} (\bibinfo{date}{2012}).
\bibitem{NaturePhysicsInsight2012blatt}
\bibinfo{author}{R.~Blatt} and \bibinfo{author}{C.~F. Roos},
  \bibinfo{title}{\emph{Quantum simulations with trapped ions}},
  \bibinfo{journal}{\Jnatphys} \bibinfo{volume}{\textbf{8}},
  \bibinfo{pages}{277} (\bibinfo{date}{2012}).
\bibitem{NaturePhysicsInsight2012aspuru-guzik}
\bibinfo{author}{A.~Aspuru-Guzik} and \bibinfo{author}{P.~Walther},
  \bibinfo{title}{\emph{Photonic quantum simulators}},
  \bibinfo{journal}{\Jnatphys} \bibinfo{volume}{\textbf{8}},
  \bibinfo{pages}{285} (\bibinfo{date}{2012}).
\bibitem{NaturePhysicsInsight2012houck}
\bibinfo{author}{A.~A. Houck}, \bibinfo{author}{H.~E. Tureci}, and
  \bibinfo{author}{J.~Koch}, \bibinfo{title}{\emph{On-chip quantum simulation
  with superconducting circuits}}, \bibinfo{journal}{\Jnatphys}
  \bibinfo{volume}{\textbf{8}}, \bibinfo{pages}{292} (\bibinfo{date}{2012}).
\bibitem{gross2017}
\bibinfo{author}{C.~Gross} and \bibinfo{author}{I.~Bloch},
  \bibinfo{title}{\emph{Quantum simulations with ultracold atoms in optical
  lattices}}, \bibinfo{journal}{\Jscience} \bibinfo{volume}{\textbf{357}},
  \bibinfo{pages}{995} (\bibinfo{date}{2017}).
\bibitem{lsp2018}
\bibinfo{author}{L.~Sanchez-Palencia}, \bibinfo{title}{\emph{Quantum
  simulation: {F}rom basic principles to applications}},
  \bibinfo{journal}{\crasphy} \bibinfo{volume}{\textbf{19}},
  \bibinfo{pages}{357} (\bibinfo{date}{2018}).
\bibitem{tarruell2018}
\bibinfo{author}{L.~Tarruell} and \bibinfo{author}{L.~Sanchez-Palencia},
  \bibinfo{title}{\emph{Quantum simulation of the {H}ubbard model with
  ultracold fermions in optical lattices}}, \bibinfo{journal}{\crasphy}
  \bibinfo{volume}{\textbf{19}}, \bibinfo{pages}{365} (\bibinfo{date}{2018}).
\bibitem{aidelsburger2018}
\bibinfo{author}{M.~Aidelsburger}, \bibinfo{author}{S.~Nascimbene}, and
  \bibinfo{author}{N.~Goldman}, \bibinfo{title}{\emph{Artificial gauge fields
  in materials and engineered systems}}, \bibinfo{journal}{\crasphy}
  \bibinfo{volume}{\textbf{19}}, \bibinfo{pages}{394} (\bibinfo{date}{2018}).
\bibitem{lebreuilly2018}
\bibinfo{author}{J.~Lebreuilly} and \bibinfo{author}{I.~Carusotto},
  \bibinfo{title}{\emph{Quantum simulation of zero temperature quantum phases
  and incompressible states of light via non-{M}arkovian reservoir engineering
  techniques}}, \bibinfo{journal}{\crasphy} \bibinfo{volume}{\textbf{19}},
  \bibinfo{pages}{433} (\bibinfo{date}{2018}).
\bibitem{LeHur2018}
\bibinfo{author}{K.~Le~Hur}, \bibinfo{author}{L.~Henriet},
  \bibinfo{author}{L.~Herviou}, \bibinfo{author}{K.~Plekhanov},
  \bibinfo{author}{A.~Petrescu}, \bibinfo{author}{T.~Goren},
  \bibinfo{author}{M.~Schiro}, \bibinfo{author}{C.~Mora}, and
  \bibinfo{author}{P.~P. Orth}, \bibinfo{title}{\emph{Driven dissipative
  dynamics and topology of quantum impurity systems}},
  \bibinfo{journal}{\crasphy} \bibinfo{volume}{\textbf{19}},
  \bibinfo{pages}{451} (\bibinfo{date}{2018}).
\bibitem{bell2018}
\bibinfo{author}{M.~Bell}, \bibinfo{author}{B.~Dou{\c{c}}ot},
  \bibinfo{author}{M.~Gershenson}, \bibinfo{author}{L.~Ioffe}, and
  \bibinfo{author}{A.~Petkovic}, \bibinfo{title}{\emph{Josephson ladders as a
  model system for 1d quantum phase transitions}}, \bibinfo{journal}{\crasphy}
  \bibinfo{volume}{\textbf{19}}, \bibinfo{pages}{484} (\bibinfo{date}{2018}).
\bibitem{alet2018}
\bibinfo{author}{F.~Alet} and \bibinfo{author}{N.~Laflorencie},
  \bibinfo{title}{\emph{Many-body localization: {A}n introduction and selected
  topics}}, \bibinfo{journal}{\crasphy} \bibinfo{volume}{\textbf{19}},
  \bibinfo{pages}{498} (\bibinfo{date}{2018}).
\bibitem{lang2018}
\bibinfo{author}{J.~Lang}, \bibinfo{author}{B.~Frank}, and
  \bibinfo{author}{J.~C. Halimeh}, \bibinfo{title}{\emph{Concurrence of
  dynamical phase transitions at finite temperature in the fully connected
  transverse-field {I}sing model}}, \bibinfo{journal}{\Jprb}
  \bibinfo{volume}{\textbf{97}}, \bibinfo{pages}{174401}
  (\bibinfo{date}{2018}).
\bibitem{halimeh2018quasiparticle}
\bibinfo{author}{J.~C. Halimeh}, \bibinfo{author}{M.~Van~Damme},
  \bibinfo{author}{V.~Zauner-Stauber}, and \bibinfo{author}{L.~Vanderstraeten},
  \bibinfo{title}{\emph{Quasiparticle origin of dynamical quantum phase
  transitions}}, \bibinfo{journal}{ar{X}iv:1810.07187}  (\bibinfo{date}{2018}).
\bibitem{hashizume2018dynamical}
\bibinfo{author}{T.~Hashizume}, \bibinfo{author}{I.~P. McCulloch}, and
  \bibinfo{author}{J.~C. Halimeh}, \bibinfo{title}{\emph{Dynamical phase
  transitions in the two-dimensional transverse-field {I}sing model}},
  \bibinfo{journal}{ar{X}iv:1811.09275}  (\bibinfo{date}{2018}).
\bibitem{prosen2000}
\bibinfo{author}{T.~Prosen}, \bibinfo{title}{\emph{Exact time-correlation
  functions of quantum {I}sing chain in a kicking transversal magnetic field:
  {S}pectral analysis of the adjoint propagator in {H}eisenberg picture}},
  \bibinfo{journal}{\Jprogthphyssup} \bibinfo{volume}{\textbf{139}},
  \bibinfo{pages}{191} (\bibinfo{date}{2000}).
\bibitem{roy2017}
\bibinfo{author}{S.~Roy}, \bibinfo{author}{R.~Moessner}, and
  \bibinfo{author}{A.~Das}, \bibinfo{title}{\emph{Locating topological phase
  transitions using nonequilibrium signatures in local bulk observables}},
  \bibinfo{journal}{\Jprb} \bibinfo{volume}{\textbf{95}},
  \bibinfo{pages}{041105} (\bibinfo{date}{2017}).
\bibitem{heyl2018}
\bibinfo{author}{M.~Heyl}, \bibinfo{author}{F.~Pollmann}, and
  \bibinfo{author}{B.~D\'ora}, \bibinfo{title}{\emph{Detecting equilibrium and
  dynamical quantum phase transitions in {I}sing chains via out-of-time-ordered
  correlators}}, \bibinfo{journal}{\Jprl} \bibinfo{volume}{\textbf{121}},
  \bibinfo{pages}{016801} (\bibinfo{date}{2018}).
\bibitem{titum2019}
\bibinfo{author}{P.~Titum}, \bibinfo{author}{J.~T. Iosue},
  \bibinfo{author}{J.~R. Garrison}, \bibinfo{author}{A.~V. Gorshkov}, and
  \bibinfo{author}{Z.-X. Gong}, \bibinfo{title}{\emph{Probing ground-state
  phase transitions through quench dynamics}},
  \bibinfo{journal}{ar{X}iv:1809.06377}  (\bibinfo{date}{2018}).
\bibitem{daug2019detection}
\bibinfo{author}{C.~B. Da\ifmmode~\breve{g}\else \u{g}\fi{}},
  \bibinfo{author}{K.~Sun}, and \bibinfo{author}{L.-M. Duan},
  \bibinfo{title}{\emph{Detection of quantum phases via out-of-time-order
  correlators}}, \bibinfo{journal}{Phys. Rev. Lett.}
  \bibinfo{volume}{\textbf{123}}, \bibinfo{pages}{140602}
  (\bibinfo{date}{2019}).
\bibitem{lieb1972finite}
\bibinfo{author}{E.~H. Lieb} and \bibinfo{author}{D.~W. Robinson},
  \bibinfo{title}{\emph{The finite group velocity of quantum spin systems}},
  \bibinfo{journal}{Comm. Math. Phys.} \bibinfo{volume}{\textbf{28}}(3),
  \bibinfo{pages}{251} (\bibinfo{date}{1972}).
\bibitem{calabrese2005evolution}
\bibinfo{author}{P.~Calabrese} and \bibinfo{author}{J.~Cardy},
  \bibinfo{title}{\emph{Evolution of entanglement entropy in one-dimensional
  systems}}, \bibinfo{journal}{J. Stat. Mech.}
  \bibinfo{volume}{\textbf{2005}}(04), \bibinfo{pages}{P04010}
  (\bibinfo{date}{2005}).
\bibitem{calabrese2006time}
\bibinfo{author}{P.~Calabrese} and \bibinfo{author}{J.~Cardy},
  \bibinfo{title}{\emph{Time dependence of correlation functions following a
  quantum quench}}, \bibinfo{journal}{Phys. Rev. Lett.}
  \bibinfo{volume}{\textbf{96}}(13), \bibinfo{pages}{136801}
  (\bibinfo{date}{2006}).
\bibitem{cevolani2018universal}
\bibinfo{author}{L.~Cevolani}, \bibinfo{author}{J.~Despres},
  \bibinfo{author}{G.~Carleo}, \bibinfo{author}{L.~Tagliacozzo}, and
  \bibinfo{author}{L.~Sanchez-Palencia}, \bibinfo{title}{\emph{Universal
  scaling laws for correlation spreading in quantum systems with short-and
  long-range interactions}}, \bibinfo{journal}{Phys. Rev. B}
  \bibinfo{volume}{\textbf{98}}(2), \bibinfo{pages}{024302}
  (\bibinfo{date}{2018}).
\bibitem{despres2019twofold}
\bibinfo{author}{J.~Despres}, \bibinfo{author}{L.~Villa}, and
  \bibinfo{author}{L.~Sanchez-Palencia}, \bibinfo{title}{\emph{Twofold
  correlation spreading in a strongly correlated lattice {B}ose gas}},
  \bibinfo{journal}{Sci. Rep.} \bibinfo{volume}{\textbf{9}}(1),
  \bibinfo{pages}{4135} (\bibinfo{date}{2019}).
\bibitem{schemmer2018}
\bibinfo{author}{M.~Schemmer}, \bibinfo{author}{A.~Johnson}, and
  \bibinfo{author}{I.~Bouchoule}, \bibinfo{title}{\emph{Monitoring squeezed
  collective modes of a one-dimensional {B}ose gas after an interaction quench
  using density-ripple analysis}}, \bibinfo{journal}{\Jpra}
  \bibinfo{volume}{\textbf{98}}, \bibinfo{pages}{043604}
  (\bibinfo{date}{2018}).
\bibitem{menu2018}
\bibinfo{author}{R.~Menu} and \bibinfo{author}{T.~Roscilde},
  \bibinfo{title}{\emph{Quench dynamics of quantum spin models with flat bands
  of excitations}}, \bibinfo{journal}{\Jprb} \bibinfo{volume}{\textbf{98}},
  \bibinfo{pages}{205145} (\bibinfo{date}{2018}).
\bibitem{fisher1989boson}
\bibinfo{author}{M.~P.~A. Fisher}, \bibinfo{author}{P.~B. Weichman},
  \bibinfo{author}{G.~Grinstein}, and \bibinfo{author}{D.~S. Fisher},
  \bibinfo{title}{\emph{Boson localization and the superfluid-insulator
  transition}}, \bibinfo{journal}{Phys. Rev. B}
  \bibinfo{volume}{\textbf{40}}(1), \bibinfo{pages}{546}
  (\bibinfo{date}{1989}).
\bibitem{cazalilla2011one}
\bibinfo{author}{M.~A. Cazalilla}, \bibinfo{author}{R.~Citro},
  \bibinfo{author}{T.~Giamarchi}, \bibinfo{author}{E.~Orignac}, and
  \bibinfo{author}{M.~Rigol}, \bibinfo{title}{\emph{One dimensional bosons:
  from condensed matter systems to ultracold gases}}, \bibinfo{journal}{Rev.
  Mod. Phys.} \bibinfo{volume}{\textbf{83}}(4), \bibinfo{pages}{1405}
  (\bibinfo{date}{2011}).
\bibitem{schutzhold2006sweeping}
\bibinfo{author}{R.~Sch{\"u}tzhold}, \bibinfo{author}{M.~Uhlmann},
  \bibinfo{author}{Y.~Xu}, and \bibinfo{author}{U.~R. Fischer},
  \bibinfo{title}{\emph{Sweeping from the superfluid to the {M}ott phase in the
  {B}ose-{H}ubbard model}}, \bibinfo{journal}{Phys. Rev. Lett.}
  \bibinfo{volume}{\textbf{97}}(20), \bibinfo{pages}{200601}
  (\bibinfo{date}{2006}).
\bibitem{fischer2008bogoliubov}
\bibinfo{author}{U.~R. Fischer}, \bibinfo{author}{R.~Sch{\"u}tzhold}, and
  \bibinfo{author}{M.~Uhlmann}, \bibinfo{title}{\emph{Bogoliubov theory of
  quantum correlations in the time-dependent {B}ose-{H}ubbard model}},
  \bibinfo{journal}{Phys. Rev. A} \bibinfo{volume}{\textbf{77}}(4),
  \bibinfo{pages}{043615} (\bibinfo{date}{2008}).
\bibitem{trotzky2012probing}
\bibinfo{author}{S.~Trotzky}, \bibinfo{author}{Y.-A. Chen},
  \bibinfo{author}{A.~Flesch}, \bibinfo{author}{I.~P. McCulloch},
  \bibinfo{author}{U.~Schollw{\"o}ck}, \bibinfo{author}{J.~Eisert}, and
  \bibinfo{author}{I.~Bloch}, \bibinfo{title}{\emph{Probing the relaxation
  towards equilibrium in an isolated strongly correlated one-dimensional {B}ose
  gas}}, \bibinfo{journal}{Nat. Phys.} \bibinfo{volume}{\textbf{8}}(4),
  \bibinfo{pages}{325} (\bibinfo{date}{2012}).
\bibitem{cheneau2012light}
\bibinfo{author}{M.~Cheneau}, \bibinfo{author}{P.~Barmettler},
  \bibinfo{author}{D.~Poletti}, \bibinfo{author}{M.~Endres},
  \bibinfo{author}{P.~Schau{\ss}}, \bibinfo{author}{T.~Fukuhara},
  \bibinfo{author}{C.~Gross}, \bibinfo{author}{I.~Bloch},
  \bibinfo{author}{C.~Kollath}, and \bibinfo{author}{S.~Kuhr},
  \bibinfo{title}{\emph{Light-cone-like spreading of correlations in a quantum
  many-body system}}, \bibinfo{journal}{Nature}
  \bibinfo{volume}{\textbf{481}}(7382), \bibinfo{pages}{484}
  (\bibinfo{date}{2012}).
\bibitem{barmettler2012propagation}
\bibinfo{author}{P.~Barmettler}, \bibinfo{author}{D.~Poletti},
  \bibinfo{author}{M.~Cheneau}, and \bibinfo{author}{C.~Kollath},
  \bibinfo{title}{\emph{Propagation front of correlations in an interacting
  {B}ose gas}}, \bibinfo{journal}{Phys. Rev. A}
  \bibinfo{volume}{\textbf{85}}(5), \bibinfo{pages}{053625}
  (\bibinfo{date}{2012}).
\bibitem{carleo2014light}
\bibinfo{author}{G.~Carleo}, \bibinfo{author}{F.~Becca},
  \bibinfo{author}{L.~Sanchez-Palencia}, \bibinfo{author}{S.~Sorella}, and
  \bibinfo{author}{M.~Fabrizio}, \bibinfo{title}{\emph{Light-cone effect and
  supersonic correlations in one-and two-dimensional bosonic superfluids}},
  \bibinfo{journal}{Phys. Rev. A} \bibinfo{volume}{\textbf{89}}(3),
  \bibinfo{pages}{031602(R)} (\bibinfo{date}{2014}).
\bibitem{villa2018cavity}
\bibinfo{author}{L.~Villa} and \bibinfo{author}{G.~De~Chiara},
  \bibinfo{title}{\emph{Cavity assisted measurements of heat and work in
  optical lattices}}, \bibinfo{journal}{Quantum} \bibinfo{volume}{\textbf{2}},
  \bibinfo{pages}{42} (\bibinfo{date}{2018}).
\bibitem{kashurnikov1996exact}
\bibinfo{author}{V.~A. Kashurnikov} and \bibinfo{author}{B.~V. Svistunov},
  \bibinfo{title}{\emph{Exact diagonalization plus renormalization-group
  theory: accurate method for a one-dimensional superfluid-insulator-transition
  study}}, \bibinfo{journal}{Phys. Rev. B} \bibinfo{volume}{\textbf{53}}(17),
  \bibinfo{pages}{11776} (\bibinfo{date}{1996}).
\bibitem{kuhner2000one}
\bibinfo{author}{T.~D. K{\"u}hner}, \bibinfo{author}{S.~R. White}, and
  \bibinfo{author}{H.~Monien}, \bibinfo{title}{\emph{One-dimensional
  {B}ose-{H}ubbard model with nearest-neighbor interaction}},
  \bibinfo{journal}{Phys. Rev. B} \bibinfo{volume}{\textbf{61}}(18),
  \bibinfo{pages}{12474} (\bibinfo{date}{2000}).
\bibitem{ejima2011dynamic}
\bibinfo{author}{S.~Ejima}, \bibinfo{author}{H.~Fehske}, and
  \bibinfo{author}{F.~Gebhard}, \bibinfo{title}{\emph{Dynamic properties of the
  one-dimensional {B}ose-{H}ubbard model}}, \bibinfo{journal}{Europhys. Lett.}
  \bibinfo{volume}{\textbf{93}}(3), \bibinfo{pages}{30002}
  (\bibinfo{date}{2011}).
\bibitem{boeris2016mott}
\bibinfo{author}{G.~Bo{\'e}ris}, \bibinfo{author}{L.~Gori},
  \bibinfo{author}{M.~D. Hoogerland}, \bibinfo{author}{A.~Kumar},
  \bibinfo{author}{E.~Lucioni}, \bibinfo{author}{L.~Tanzi},
  \bibinfo{author}{M.~Inguscio}, \bibinfo{author}{T.~Giamarchi},
  \bibinfo{author}{C.~D'Errico}, \bibinfo{author}{G.~Carleo}, \emph{et~al.},
  \bibinfo{title}{\emph{Mott transition for strongly interacting
  one-dimensional bosons in a shallow periodic potential}},
  \bibinfo{journal}{Phys. Rev. A} \bibinfo{volume}{\textbf{93}}(1),
  \bibinfo{pages}{011601(R)} (\bibinfo{date}{2016}).
\bibitem{kohlert2019observation}
\bibinfo{author}{T.~Kohlert}, \bibinfo{author}{S.~Scherg},
  \bibinfo{author}{X.~Li}, \bibinfo{author}{H.~P. L{\"u}schen},
  \bibinfo{author}{S.~DasSarma}, \bibinfo{author}{I.~Bloch}, and
  \bibinfo{author}{M.~Aidelsburger}, \bibinfo{title}{\emph{Observation of
  many-body localization in a one-dimensional system with a single-particle
  mobility edge}}, \bibinfo{journal}{Phys. Rev. Lett.}
  \bibinfo{volume}{\textbf{122}}(17), \bibinfo{pages}{170403}
  (\bibinfo{date}{2019}).
\bibitem{pitaevskii2004}
\bibinfo{author}{L.~P. Pitaevskii} and \bibinfo{author}{S.~Stringari},
  \bibinfo{title}{\emph{{{B}ose-{E}instein {C}ondensation}}}
  (\bibinfo{publisher}{Clarendon Press, Oxford}, \bibinfo{year}{2004}).
\bibitem{lieb1963exact1}
\bibinfo{author}{E.~H. Lieb} and \bibinfo{author}{W.~Liniger},
  \bibinfo{title}{\emph{Exact analysis of an interacting {B}ose gas. {I}. the
  general solution and the ground state}}, \bibinfo{journal}{Phys. Rev.}
  \bibinfo{volume}{\textbf{130}}(4), \bibinfo{pages}{1605}
  (\bibinfo{date}{1963}).
\bibitem{lieb1963exact2}
\bibinfo{author}{E.~H. Lieb}, \bibinfo{title}{\emph{Exact analysis of an
  interacting {B}ose gas. {II}. the excitation spectrum}},
  \bibinfo{journal}{Phys. Rev.} \bibinfo{volume}{\textbf{130}}(4),
  \bibinfo{pages}{1616} (\bibinfo{date}{1963}).
\bibitem{haldane1980solidification}
\bibinfo{author}{F.~Haldane}, \bibinfo{title}{\emph{Solidification in a soluble
  model of bosons on a one-dimensional lattice: the boson-{H}ubbard chain}},
  \bibinfo{journal}{Phys. Lett. A} \bibinfo{volume}{\textbf{80}}(4),
  \bibinfo{pages}{281} (\bibinfo{date}{1980}).
\bibitem{krauth1991bethe}
\bibinfo{author}{W.~Krauth}, \bibinfo{title}{\emph{Bethe ansatz for the
  one-dimensional boson {H}ubbard model}}, \bibinfo{journal}{Phys. Rev. B}
  \bibinfo{volume}{\textbf{44}}(17), \bibinfo{pages}{9772}
  (\bibinfo{date}{1991}).
\bibitem{kiwata1994bethe}
\bibinfo{author}{H.~Kiwata} and \bibinfo{author}{Y.~Akutsu},
  \bibinfo{title}{\emph{Bethe-ansatz approximation for the {S}=1
  antiferromagnetic spin chain}}, \bibinfo{journal}{J. Phys. Soc. Japan}
  \bibinfo{volume}{\textbf{63}}(10), \bibinfo{pages}{3598}
  (\bibinfo{date}{1994}).
\bibitem{choy1982failure}
\bibinfo{author}{T.~Choy} and \bibinfo{author}{F.~Haldane},
  \bibinfo{title}{\emph{Failure of {B}ethe-ansatz solutions of generalisations
  of the {H}ubbard chain to arbitrary permutation symmetry}},
  \bibinfo{journal}{Phys. Lett. A} \bibinfo{volume}{\textbf{90}}(1-2),
  \bibinfo{pages}{83} (\bibinfo{date}{1982}).
\bibitem{note:exp_triple_occupancy}
Triple occupancy has been shown to be vanishingly small for the BHm and
  interaction strengths above $U/J \simeq 20$, see for instance Fig.~S3 in the
  Supplemental of \cite{ronzheimer2013expansion}.
\bibitem{kashurnikov1998zero}
\bibinfo{author}{V.~A. Kashurnikov}, \bibinfo{author}{A.~V. Krasavin}, and
  \bibinfo{author}{B.~V. Svistunov}, \bibinfo{title}{\emph{Zero-point phase
  transitions in the one-dimensional truncated bosonic {H}ubbard model and its
  spin-1 analog}}, \bibinfo{journal}{Phys. Rev. B}
  \bibinfo{volume}{\textbf{58}}(4), \bibinfo{pages}{1826}
  (\bibinfo{date}{1998}).
\bibitem{cazalilla2004differences}
\bibinfo{author}{M.~A. Cazalilla}, \bibinfo{title}{\emph{Differences between
  the {T}onks regimes in the continuum and on the lattice}},
  \bibinfo{journal}{Phys. Rev. A} \bibinfo{volume}{\textbf{70}}(4),
  \bibinfo{pages}{041604(R)} (\bibinfo{date}{2004}).
\bibitem{note:DoubleStructure}
We have checked that, for any value of the filling, the QSF computed here is
  compatible with the characteristic velocities of the correlation spreading
  reported in Ref~\cite{despres2019twofold}. We confirm that the correlation
  edge velocity is twice the maximal group velocity, while the local
  correlation maxima propagate at twice the corresponding phase velocity.
\bibitem{jurcevic2014}
\bibinfo{author}{P.~Jurcevic}, \bibinfo{author}{B.~P. Lanyon},
  \bibinfo{author}{P.~Hauke}, \bibinfo{author}{C.~Hempel},
  \bibinfo{author}{P.~Zoller}, \bibinfo{author}{R.~Blatt}, and
  \bibinfo{author}{C.~F. Roos}, \bibinfo{title}{\emph{Quasiparticle engineering
  and entanglement propagation in a quantum many-body system}},
  \bibinfo{journal}{\Jnature} \bibinfo{volume}{\textbf{511}},
  \bibinfo{pages}{202} (\bibinfo{date}{2014}).
\bibitem{richerme2014}
\bibinfo{author}{P.~Richerme}, \bibinfo{author}{Z.-X. Gong},
  \bibinfo{author}{A.~Lee}, \bibinfo{author}{C.~Senko},
  \bibinfo{author}{J.~Smith}, \bibinfo{author}{M.~Foss-Feig},
  \bibinfo{author}{S.~Michalakis}, \bibinfo{author}{A.~V. Gorshkov}, and
  \bibinfo{author}{C.~Monroe}, \bibinfo{title}{\emph{Non-local propagation of
  correlations in quantum systems with long-range interactions}},
  \bibinfo{journal}{\Jnature} \bibinfo{volume}{\textbf{511}},
  \bibinfo{pages}{198} (\bibinfo{date}{2014}).
\bibitem{hauke2013spread}
\bibinfo{author}{P.~Hauke} and \bibinfo{author}{L.~Tagliacozzo},
  \bibinfo{title}{\emph{Spread of correlations in long-range interacting
  quantum systems}}, \bibinfo{journal}{Phys. Rev. Lett.}
  \bibinfo{volume}{\textbf{111}}(20), \bibinfo{pages}{207202}
  (\bibinfo{date}{2013}).
\bibitem{schachenmayer2013}
\bibinfo{author}{J.~Schachenmayer}, \bibinfo{author}{B.~P. Lanyon},
  \bibinfo{author}{C.~F. Roos}, and \bibinfo{author}{A.~J. Daley},
  \bibinfo{title}{\emph{Entanglement growth in quench dynamics with variable
  range interactions}}, \bibinfo{journal}{\Jprx} \bibinfo{volume}{\textbf{3}},
  \bibinfo{pages}{031015} (\bibinfo{date}{2013}).
\bibitem{cevolani2015protected}
\bibinfo{author}{L.~Cevolani}, \bibinfo{author}{G.~Carleo}, and
  \bibinfo{author}{L.~Sanchez-Palencia}, \bibinfo{title}{\emph{Protected
  quasilocality in quantum systems with long-range interactions}},
  \bibinfo{journal}{Phys. Rev. A} \bibinfo{volume}{\textbf{92}}(4),
  \bibinfo{pages}{041603(R)} (\bibinfo{date}{2015}).
\bibitem{cevolani2016spreading}
\bibinfo{author}{L.~Cevolani}, \bibinfo{author}{G.~Carleo}, and
  \bibinfo{author}{L.~Sanchez-Palencia}, \bibinfo{title}{\emph{Spreading of
  correlations in exactly solvable quantum models with long-range interactions
  in arbitrary dimensions}}, \bibinfo{journal}{New J. Phys.}
  \bibinfo{volume}{\textbf{18}}(9), \bibinfo{pages}{093002}
  (\bibinfo{date}{2016}).
\bibitem{buyskikh2016}
\bibinfo{author}{A.~S. Buyskikh}, \bibinfo{author}{M.~Fagotti},
  \bibinfo{author}{J.~Schachenmayer}, \bibinfo{author}{F.~Essler}, and
  \bibinfo{author}{A.~J. Daley}, \bibinfo{title}{\emph{Entanglement growth and
  correlation spreading with variable-range interactions in spin and fermionic
  tunneling models}}, \bibinfo{journal}{\Jpra} \bibinfo{volume}{\textbf{93}},
  \bibinfo{pages}{053620} (\bibinfo{date}{2016}).
\bibitem{koffel2012entanglement}
\bibinfo{author}{T.~Koffel}, \bibinfo{author}{M.~Lewenstein}, and
  \bibinfo{author}{L.~Tagliacozzo}, \bibinfo{title}{\emph{Entanglement entropy
  for the long-range {I}sing chain in a transverse field}},
  \bibinfo{journal}{Phys. Rev. Lett.} \bibinfo{volume}{\textbf{109}}(26),
  \bibinfo{pages}{267203} (\bibinfo{date}{2012}).
\bibitem{dolfi2014matrix}
\bibinfo{author}{M.~Dolfi}, \bibinfo{author}{B.~Bauer},
  \bibinfo{author}{S.~Keller}, \bibinfo{author}{A.~Kosenkov},
  \bibinfo{author}{T.~Ewart}, \bibinfo{author}{A.~Kantian},
  \bibinfo{author}{T.~Giamarchi}, and \bibinfo{author}{M.~Troyer},
  \bibinfo{title}{\emph{Matrix product state applications for the {ALPS}
  project}}, \bibinfo{journal}{Comput. Phys. Commun.}
  \bibinfo{volume}{\textbf{185}}(12), \bibinfo{pages}{3430}
  (\bibinfo{date}{2014}).
\bibitem{popov1983}
\bibinfo{author}{V.~M. Popov}, \bibinfo{title}{\emph{{Functional Integrals in
  Quantum Field Theory and Statistical Physics}}} (\bibinfo{publisher}{Reidel,
  Dordrecht}, \bibinfo{year}{1983}).
\bibitem{mora2003}
\bibinfo{author}{C.~Mora} and \bibinfo{author}{Y.~Castin},
  \bibinfo{title}{\emph{Extension of {B}ogoliubov theory to quasicondensates}},
  \bibinfo{journal}{\Jpra} \bibinfo{volume}{\textbf{67}}(5),
  \bibinfo{pages}{053615} (\bibinfo{date}{2003}).
\bibitem{holstein1940}
\bibinfo{author}{T.~Holstein} and \bibinfo{author}{H.~Primakoff},
  \bibinfo{title}{\emph{Field dependence of the intrinsic domain magnetization
  of a ferromagnet}}, \bibinfo{journal}{\Jpr} \bibinfo{volume}{\textbf{58}},
  \bibinfo{pages}{1098} (\bibinfo{date}{1940}).
\bibitem{auerbach1994}
\bibinfo{author}{A.~Auerbach}, \bibinfo{title}{\emph{{Interacting Electrons and
  Quantum Magnetism}}} (\bibinfo{publisher}{Springer, {N}ew {Y}ork},
  \bibinfo{year}{1994}).
\bibitem{korepin1997quantum}
\bibinfo{author}{V.~E. Korepin}, \bibinfo{author}{N.~M. Bogoliubov}, and
  \bibinfo{author}{A.~G. Izergin}, \bibinfo{title}{\emph{Quantum Inverse
  Scattering Method and Correlation Functions}}, \bibinfo{volume}{vol.~3}
  (\bibinfo{publisher}{Cambridge University Press}, \bibinfo{year}{1997}).
\bibitem{sutherland2004beautiful}
\bibinfo{author}{B.~Sutherland}, \bibinfo{title}{\emph{Beautiful Models: 70
  Years of Exactly Solved Quantum Many-Body Problems}}
  (\bibinfo{publisher}{World Scientific Publishing Company, Singapore},
  \bibinfo{year}{2004}).
\bibitem{franchini2017introduction}
\bibinfo{author}{F.~Franchini}, \bibinfo{title}{\emph{An Introduction to
  Integrable Techniques for One-Dimensional Quantum Systems}},
  \bibinfo{volume}{vol. 940} (\bibinfo{publisher}{Springer},
  \bibinfo{year}{2017}).
\bibitem{ronzheimer2013expansion}
\bibinfo{author}{J.~P. Ronzheimer}, \bibinfo{author}{M.~Schreiber},
  \bibinfo{author}{S.~Braun}, \bibinfo{author}{S.~S. Hodgman},
  \bibinfo{author}{S.~Langer}, \bibinfo{author}{I.~P. McCulloch},
  \bibinfo{author}{F.~Heidrich-Meisner}, \bibinfo{author}{I.~Bloch}, and
  \bibinfo{author}{U.~Schneider}, \bibinfo{title}{\emph{Expansion dynamics of
  interacting bosons in homogeneous lattices in one and two dimensions}},
  \bibinfo{journal}{Phys. Rev. Lett.} \bibinfo{volume}{\textbf{110}}(20),
  \bibinfo{pages}{205301} (\bibinfo{date}{2013}).
\bibitem{yang1969thermodynamics}
\bibinfo{author}{C.-N. Yang} and \bibinfo{author}{C.~P. Yang},
  \bibinfo{title}{\emph{Thermodynamics of a one-dimensional system of bosons
  with repulsive delta-function interaction}}, \bibinfo{journal}{Journal of
  Mathematical Physics} \bibinfo{volume}{\textbf{10}}(7), \bibinfo{pages}{1115}
  (\bibinfo{date}{1969}).

\end{thebibliography}

\end{document}